# Hypothetical bias in stated choice experiments: Part II. Macro-scale analysis of literature and effectiveness of bias mitigation methods

Milad Haghani[1,*], Michiel C. J. Bliemer[1], John M. Rose[2], Harmen Oppewal[3], Emily Lancsar[4]

[1] Institute of Transport and Logistics Studies, The University of Sydney Business School, The University of Sydney, Australia
[2] Centre for Business Intelligence and Data Analytics, UTS Business School, University of Technology Sydney, Australia
[3] Department of Marketing, Monash Business School, Monash University, Australia
[4] Department of Health Services Research and Policy, Research School of Population Health, Australian National University, Australia

[*]corresponding author: milad.haghani@sydney.edu.au

## Abstract

This paper reviews methods of hypothetical bias (HB) mitigation in choice experiments (CEs). It presents a bibliometric analysis and summary of empirical evidence of their effectiveness. The paper follows the review of empirical evidence on the existence of HB presented in Part I of this study. While the number of CE studies has rapidly increased since 2010, the critical issue of HB has been studied in only a small fraction of CE studies. The present review includes both ex-ante and ex-post bias mitigation methods. Ex-ante bias mitigation methods include cheap talk, real talk, consequentiality scripts, solemn oath scripts, opt-out reminders, budget reminders, honesty priming, induced truth telling, indirect questioning, time to think and pivot designs. Ex-post methods include follow-up certainty calibration scales, respondent perceived consequentiality scales, and revealed-preference-assisted estimation. It is observed that the use of mitigation methods markedly varies across different sectors of applied economics. The existing empirical evidence points to their overall effectives in reducing HB, although there is some variation. The paper further discusses how each mitigation method can counter a certain subset of HB sources. Considering the prevalence of HB in CEs and the effectiveness of bias mitigation methods, it is recommended that implementation of at least one bias mitigation method (or a suitable combination where possible) becomes standard practice in conducting CEs. Mitigation method(s) suited to the particular application should be implemented to ensure that inferences and subsequent policy decisions are as much as possible free of HB.

## 1. Introduction

The debate on the relevance of experimental data in economic valuation dates as far back as two centuries ago (Ben-Akiva et al., 2019; Fisher, 1893). From its inception, the notion of *stated preference* (SP) (Thurstone, 1931), has sparked debate as to whether consumer experiments should go anywhere beyond the mere testing of axioms of economic choice and whether SP data collected outside of real-world markets can be utilised for demand estimation (Ben-Akiva et al., 2019). It was not until the 1960's (Luce, 1965; Luce and Tukey, 1964) that sustained application of SP methods in the form of ratings- and *choice-based conjoint* measurement began to emerge in the academic literature (Green, 1974; Green and Srinivasan, 1978; Johnson, 1974; Louviere and Woodworth, 1983). Since then, choice-based SP methods have increasingly become established as a well-founded and preferred approach of studying choice behaviour (Louviere et al., 2000). They are also increasingly preferred over *contingent valuation* (CV) (Ciriacy-Wantrup, 1947; Davis, 1963), a method that environmental economists originally proposed to determine monetary values of environmental damage/protection (Bergstrom, 1990; Thayer, 1981). *Choice experiments* (CEs), also known by a variety of other terms, including *stated choice experiments* and *discrete choice experiments*, represent an indirect approach of inferring preferences. CEs present subjects with hypothetical profiles of multiple alternatives described by their attribute levels. The subject states their preferred option and subsequent analysis reveals the subject's implied utilities of the alternatives and their attribute levels. A subject may be a potential commuter (Devarasetty et al., 2012), food consumer (Alfnes et al., 2006), energy consumer (Mamkhezri et al., 2020), patient (de Bekker-Grob et al., 2019), physician (Kulik and Carlino, 1987), farmer (Espinosa-Goded et al., 2010), air traveller (Hensher and Louviere, 1983), tourist (Oppewal et al., 2015; van Cranenburgh et al., 2014) or pollution/climate conservation tax payer (Liu et al., 2015; Ščasný et al., 2017), to name only a few of the many applications.

Choice observations are usually analysed using discrete choice analysis using the theory of random utility maximization (McFadden, 1981, 1986; McFadden et al., 1986). This provides the analyst an array of econometric measures for placing a value on non-traded goods, market shares of novel products, or price sensitivities and elasticities. The use of CEs also allows the analyst to circumvent the potential issues linked to the acquisition of choice observations from real markets; i.e., *revealed preferences* (RP) data, namely, ambiguity in the delineation of the choice set and in the attributes of non-chosen alternatives and the absence of adequate attribute variability due to forces of market equilibrium. Despite all the appealing features, the inherent lack of realism arising from the hypothetical nature of CEs, i.e., the issue of *hypothetical bias (HB)*, has always loomed over the use of CEs, meaning that their results have been viewed with scepticism (Mitani and Flores, 2014). As such, the potential presence of HB could limit, for some, the acceptance and use of these methods in economic valuation studies.

With respect to the usefulness of CEs, over the recent years much emphasis has been placed on enhancing their *statistical efficiency*, to obtain more reliable estimates (than, for example, those obtained from conventional orthogonal designs) from a given sample size (Rose and Bliemer, 2009; Rose et al., 2008), as well as avoiding dominant alternatives (Bliemer et al., 2017). Efforts to design efficient CEs however will be only effective if there is sufficient *behavioural realism* in the underlying data (Hensher, 2010, 2015). In many situations involving non-market or non-existent goods, there is no plausible alternative for CEs. However, we argue that issues surrounding behavioural realism of CEs should not be viewed as whether analytical advantages of CEs justify the lack of realism or whether there is any better alternative. Rather, actions need to be taken to improve behavioural realism and address such issues during both design and analysis phases of such surveys. It is important to consider practical implications of HB in high-stake CE-based cost-benefit analyses where an uncontrolled bias could amount to the make-or-break of major national projects. Therefore, rather than simply accepting the existence of HB as an inherent feature of CEs and hoping for the best, we argue that it is essential that those who employ CEs have a nuanced appreciation of the HB problem including its likely sources, its likely direction and magnitude and also, of ways to effectively counter/minimise bias during the experiment design/administration and/or analysis phases.

In developing this detailed understanding of HB, it is critical to maximally and rigorously test the problem in as many contexts of choice as possible to form a database that informs us about the likely prevalence of the problem and its underlying causes and drivers. And given inherent structural differences of CEs with other forms of SP (e.g., CVs), such inferences should be obtained from the evidence on choice methods rather than borrowing evidence from the CV domain. In a comprehensive synthesis of such empirical evidence in Part I of this study (Haghani et al., 2020), we established that although there is variability in findings on the existence of HB in CEs, when one considers the entirety of empirical evidence, the role of HB in CEs is undeniable.

Following up on Haghani et al. (2020), the present paper analyses the strategies for HB mitigation and the empirical evidence on their effectiveness. The aim is to disentangle the evidence specific to CEs from the evidence in the CV domain and to develop a synthesis of findings exclusive to CEs. In addition, a macro scale bibliometric analysis of the CE literature is provided to gain broader insight into the development of CE methods.

In the remainder of this paper, Section 2 describes the search strategies and datasets used for our analysis. Section 3 presents the macro scale analysis of various aspects of the CE literature. Section 4 reviews empirical evidence on HB mitigation methods. Section 5 shows how mitigation strategies can be linked to sources of hypothetical bias and how by recognising the likely sources, tailored mitigation strategies may be adopted for each CE application to minimise HB. Section 6 discusses the prevalence of mitigation methods and studies, and Section 7 presents a summary and draws final conclusions.

## 2. Data and methods

Three main keyword-based search strategies generated a dataset of empirical studies in the CE domain that have tested effectiveness of HB mitigation strategies. Using the Web of Science (WoS) Core Collection as the main search tool, the following queries were devised to search the title, abstract and keywords of the indexed studies.

Query (i): "hypothetical bias"

Query (ii): ("stated choice*" OR "stated preference*" OR "choice experiment*" OR "discrete choice*" OR ("conjoint analysis" AND "choice*") OR "conjoint choice" OR "choice based conjoint") AND ("hypothetical bias" OR "external validity")

Query (iii): ("hypothetical bias") AND ("mitigate" OR "mitigation" OR "cheap talk" OR ("certainty" AND "calibration") OR "ex-ante" OR "ex-post" OR "oath" OR "time to think" OR "time-to-think" OR "deliberation" OR "deliberate" OR "honesty")

The three searches were first conducted in November 2018 and were updated in April 2020, with respectively 429, 235 and 355 items returned during the last update. No restriction on time span or document type was imposed. Items were filtered against the set of inclusion and exclusion criteria presented below. Peer reviewed journal articles that report empirical evaluations on the effectiveness of mitigation methods were included in the *core dataset* of references. Articles studying mitigation methods of HB in experimental settings other than CE, e.g., bias mitigation experiments in CV (Lawton et al., 2019; Silva et al., 2011) or auctions (Furno et al., 2019) or referendum voting (Mozumder and Berrens, 2007), were not included. Also excluded were CE studies in which a mitigation method was adopted purely as a solution to HB without testing its effect (Amilon et al., 2020; Van Loo et al., 2011; Wuepper et al., 2019). To maximise the inclusivity of the core dataset, supplementary searches with a flexible structure were conducted in Google Scholar. In addition, we applied limited forward and backward expansion searching on initially identified items. A total of 56 articles qualified for inclusion in the core dataset of articles to be individually examined.

The core set of articles on the effectiveness of mitigation strategies firstly includes studies that compare estimates with versus without a mitigation strategy with a realistic benchmark, thereby measuring *absolute* effectiveness. They also include studies that draw comparisons between estimates obtained from two hypothetical treatments with versus without implementation of a mitigation strategy, thus measuring *relative* effectiveness.

Qualified peer-reviewed articles were examined individually to extract relevant information and conclusions. These include the type of mitigation method(s) investigated, the choice context in which the evaluation was made, whether the design was based on a between or within subject comparison, whether absolute or relative bias was measured, whether the method was effective in reducing the HB (as interpreted and reported by authors of each study), and the main highlights of each study. The effectiveness was deemed “mixed” if, for example, the mitigation strategy reduced HB based on a certain metric but was ineffective (or amplified the bias) based on another metric.

For the macro-level analysis, a supplementary query (iv), was formulated as below.

Query (iv): “stated choice*” OR (“stated preference*” AND “choice*”) OR “discrete choice experiment*” OR (“conjoint analysis” AND “choice”) OR "conjoint choice" OR "choice based conjoint"

The search returned 7,332 items. Full bibliometric details of all these articles were exported from the WoS, including citation information (author(s), document title, year of publication, source title, citation count, source and document type), bibliographic information (e.g. affiliations, abbreviated source title), abstract, keywords and list of references. These were next analysed using the scientometric software VOSviewer (Van Eck and Waltman, 2010).

## 3. Macro-scale analysis of the literature

Applications of CEs have become increasingly common across different sectors of applied economics. While there were early pioneering studies (Thurstone, 1931), the method has been gaining traction from the early 1980s and increasing usage over the 1990s. The WoS record shows less than 10 CE-related publications during the 1980s. Some of the earliest studies concern applications in consumer shopping (Louviere, 1984; Moore, 1989) and commuter travel (Fowkes and Wardman, 1988; Hensher et al., 1988). Sustained application of CEs, however, emerged since 1990, with early studies dominated by transport-related applications (Axhausen and Polak, 1991; Bunch et al., 1993; Fowkes et al., 1991; Hensher, 1994; Khattak et al., 1993; Polak and Jones, 1993; Vanderwaerden et al., 1993; Wardman, 1991). Louviere and Timmermans (1992) may be regarded as one of the first studies raising the issue of *external validity* (EV) in this domain. HB can be seen as a main component influencing EV, see Haghani et al. (2020). Figure 1 displays the number of CE-related studies published each year since 1990, along with a similar estimate for CV-related studies (based on counting the number of articles that have cited the term “contingent valuation” in their title or abstract or keyword list). Between 1990 and 2000, annually between four and 49 studies, and a total of nearly 230 CE-related studies were published according to our data. In contrast, the annual number of studies linked to CV, the other and at that time more popular SP method, varied between 16 and 139, with an estimated total of nearly 900 publications over the same time. The figure also shows the number of CE publications citing the term HB and among them those that have cited either of the terms HB or EV in their title, abstract or keywords.

These results further show how, since 2010, the number of CE studies published per year has surpassed that of their CV counterpart. 2014 shows a notable increase in the rate of CE-related publications, largely attributable to an increased attention in the health domain. In fact, it seems that since 2014, more studies linked to CEs have appeared every year in health-related journals than any of the other identified sectors of applied economics (i.e., consumer, transport and environmental/resource economics). The fraction of CE studies that have cited HB however has remained very small. While there appears to be a correspondence between the number of CE articles published each year and those related to HB/EV, the portion has almost invariably been small throughput the history of these publications. For example, in 2019 only 2.3 percent of the entire body of studies published in the CE literature (more than 900 items) have a link to HB/EV. Across all years this percentage is only 2.7 percent.

Given the centrality of the issue of HB to CE applications, the fact that studies on HB/EV make up such a small fraction of this voluminous literature may be due to empirical evaluations of HB being fundamentally challenging. One of the challenges is the lack of benchmarks to base the evaluation on (Loomis, 2011). This has essentially made the issue to remain rather a mystery and often a matter of speculation in many CE applications. As Mitani and Flores (2014) state, while "the issue of hypothetical bias in stated preference economic analysis is one of historical and ongoing concern", to date, "there has been no systematic explanation of hypothetical bias and the underlying causes have not been yet sufficiently understood" (p. 452). Or, as Loomis (2011) states "there is no widely accepted general theory of respondent behaviour that explains hypothetical bias" (p. 363).

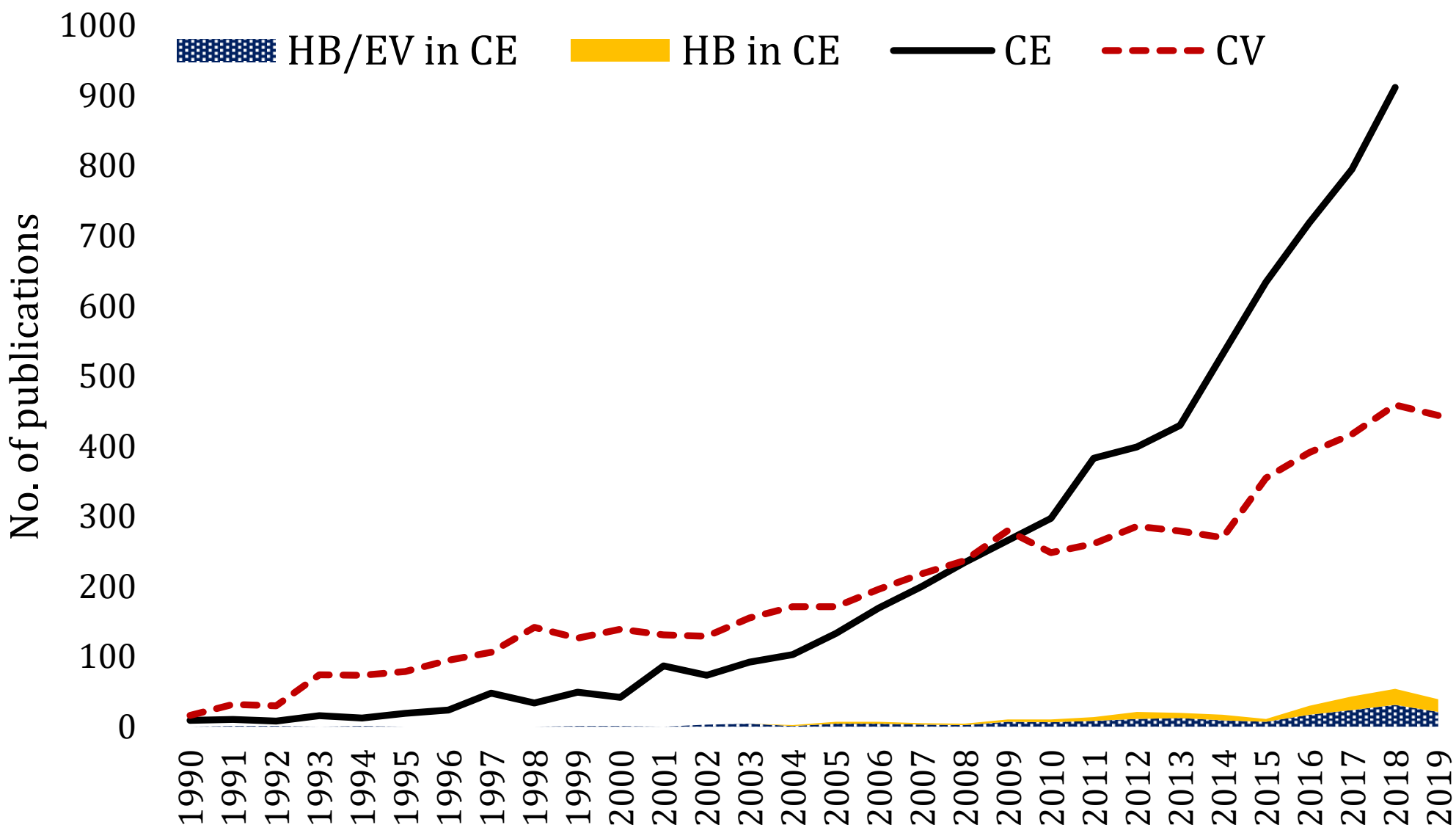

**Figure 1** Per annum number of publications linked to CE (or CV) methods, as well as those of CE linked to HB/EV.

As pointed out earlier, the four major sectors of applied economics we identify for this analysis (i.e., environmental, consumer, health and transport) have consistently used CE methods for preference valuation. Figure 2 lists the 50 journals that have published most CE articles along with the number for each. The four major areas of applied economics have been colour-coded. Figure 3 provides a different perspective to the relationships between journals by visualising their bibliographic coupling relationships, i.e., the similarity of the references of their underlying articles related to CEs. Bibliographic coupling indicates the thematic similarities between publications (Kessler, 1963). In this visualisation, each journal is represented by a node whose size is proportional to its number of publications. Journals that are closely coupled are visualised closer to each other and can form a cluster. The VOSviewer algorithm identified four major clusters of bibliographically coupled journals in the CE literature. Upon inspection, it becomes clear that these four clusters do, in fact, closely coincide with the four sectors of applied economics identified above. The colour assigned to each cluster matches the colour-coding in Figure 2.

As evident from these analyses, *Value in Health* has by far published the most CE-related studies. This is followed by *Transportation Research Part A* with more than 200 CE-related publications at the time of this analysis. The map of bibliographic coupling displays rather homogenous and well-defined clusters attributable to transport, health, and environmental studies. In contrast, the fourth cluster is rather dispersed across the map and mostly comprises consumer choice studies. The node sizes in this cluster are also smaller than those of the three other areas, meaning that of the four major areas of applied economics where CE applications have been consistently reported, consumer economics is the one with the least publications. In terms of the highest number of articles in the respective disciplines, *Ecological Economics* and *Environment and Resource*

*Economics* has been the most active outlet in the environmental domain, *Transportation Research Part A* in the transport domain, *Journal of Marketing Research*, *Food Quality and Preference* and *Marketing Science* in the consumer domain, and *Value in Health* in the health domain. The *Journal of Choice Modelling*, despite its multidisciplinary nature, is most closely identified with the transport cluster (although its position towards the centre of the map does reflect the variety of disciplines of origins of its CE publications), while *PLOS ONE*, another multidisciplinary outlet, seems to identify most closely with the health cluster.

The co-occurrence of keywords in CE publications was also analysed and the network is shown in Figure 4. Two keywords are considered as co-occurring when they are listed in the same publication. Keywords that more frequently co-occurred are more connected and appear closer to each other in the map. The size of each node is proportional to the frequency of the keyword that it represents. Again, four major clusters can be identified, matching the previous categorisation and colour-coding in Figure 2. The term "*willingness to pay*" seems to be the single most frequent keyword in CE-related articles. Discipline-specific preferences for characterising CE studies are observable through this map. While consumer economists have frequently used the term "*conjoint analysis*" to characterise their CE studies, the terms "*choice experiments*" and "*discrete choice experiment*" are shown to have been more common and popular in the health domain. Transport and environmental researchers, on the other hand, have more frequently characterised their studies in this domain as "*stated preference (survey)*". Also, the keyword "*contingent valuation*" appears to accompany many CE studies across all four domains, but particularly, in environmental applications.

The clusters give an idea of the extent and versatility of CE applications within each sector. Environmental studies have largely focused on issues such as "*nonmarket benefits*" and "*cost-benefit analysis*" of policies and projects related to "*environmental valuation*" and "*public goods*", including those linked to "*renewable energy*", "*green electricity*", "*ecosystem services*", "*climate change*" and environmental "*conservation*", as well as the estimation of "*recreation demand*". Many of the meta-analytical studies have originated from this field, and this is reflected in the map by the fact that the term "*meta-analysis*" is identified with this cluster. In terms of methodological issues that have been particularly of interest to environmental scientists, the issue of "*spatial heterogeneity*" stands out. Also, "*incentive compatibility*", which is closely related to "*hypothetical bias*", seems to have been discussed more heavily by environmental studies than any other domain.

Studies of "*consumer preference*" seem to have largely focused on consumer "*food choice*", particularly that of "*beef*", "*meat*", "*fruit*" and "*organic food*" choice, as well as issues surrounding "*marketing*" of food such as "*food labels*" and "*pricing*". The increasing applications of "*eye-tracking*" technologies in consumer choice experiments is also reflected in the keywords. Among methodological and econometric issues of choice, it seems that the issue of "*consumer heterogeneity*" has received considerable attention from researchers of this field.

The figure further shows that health studies have applied CE methods for the evaluation of "*public preferences*" for various health "*services*" as well as "*patient*" and "*doctor*" preferences and "*decision-making*" for "*drug-treatments*" and "*therapies*" for a broad range of mental and physical health problems. In transport studies, however, travel "*demand*" estimation (associated with "*public transit*", road "*networks*", "*air transport*"), as well as the (value of) "*travel time*" (saving), and "*travel-time reliability/variability*" seem to have been the centre of attention, as essential inputs for transport projects such as "*road pricing*" schemes. More contemporary topics studied in transport seem to be related to "*alternative fuel vehicles*" and "*electric vehicles*" as well as the value of and commuter reaction to "*traffic information*" and "*adoption*" of new transport technologies such as "*autonomous vehicles*" and "*high-speed rail*". The majority of methodological advancements in "*survey design*" as well as those related to capturing econometric phenomena linked to error-term specification have been discussed in conjunction with transport-related applications. The terms "*mixed logit*" and "*latent-class model*" are more closely associated with transport than with any other cluster.

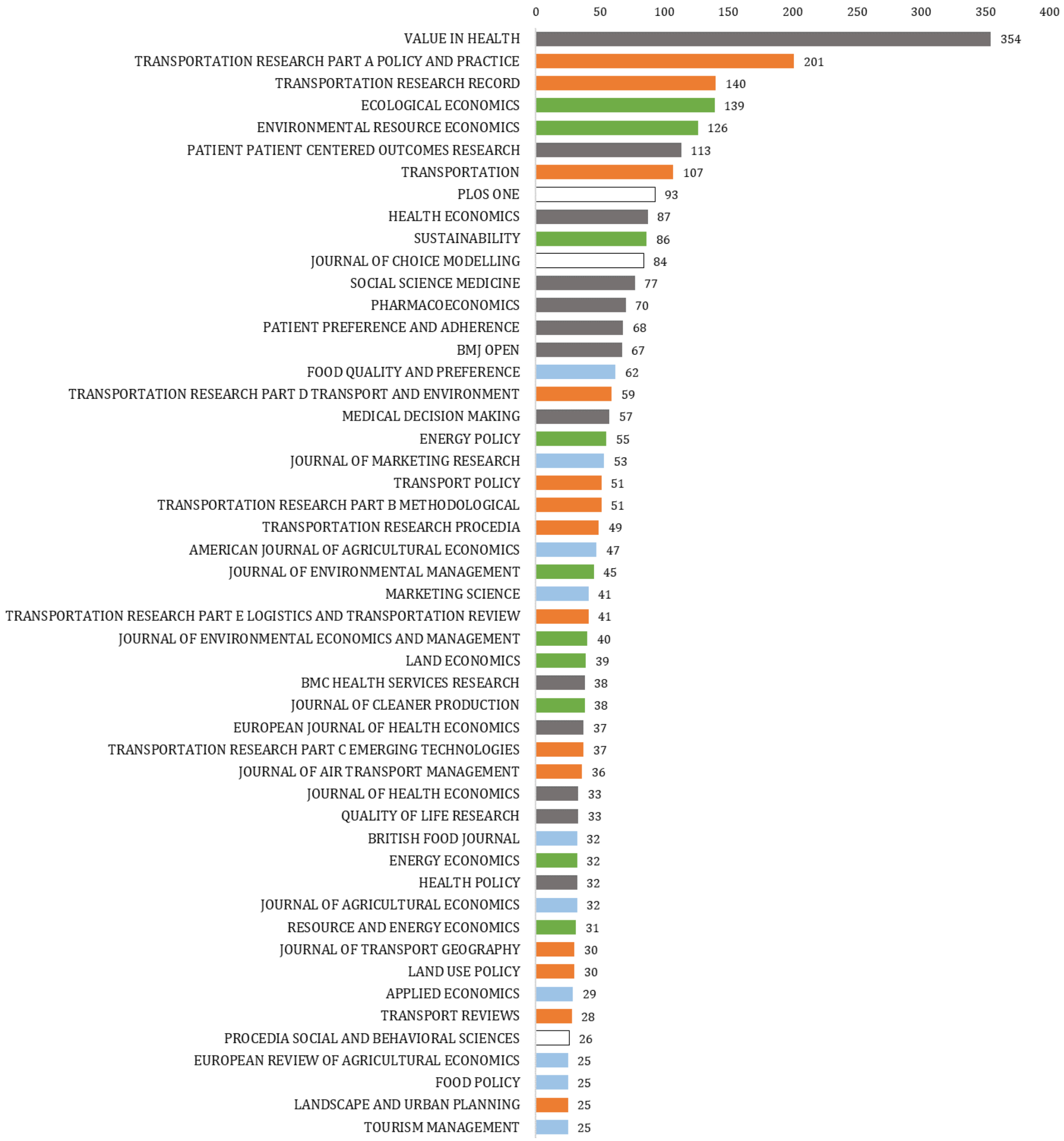

**Figure 2** Fifty journals with largest number of publications related to CEs. (Legend: grey = journals predominantly related to health applications; orange = transport; green = environmental; blue = consumer studies; white = journals whose publications are mixed and not attributable to any of the four mentioned areas.)

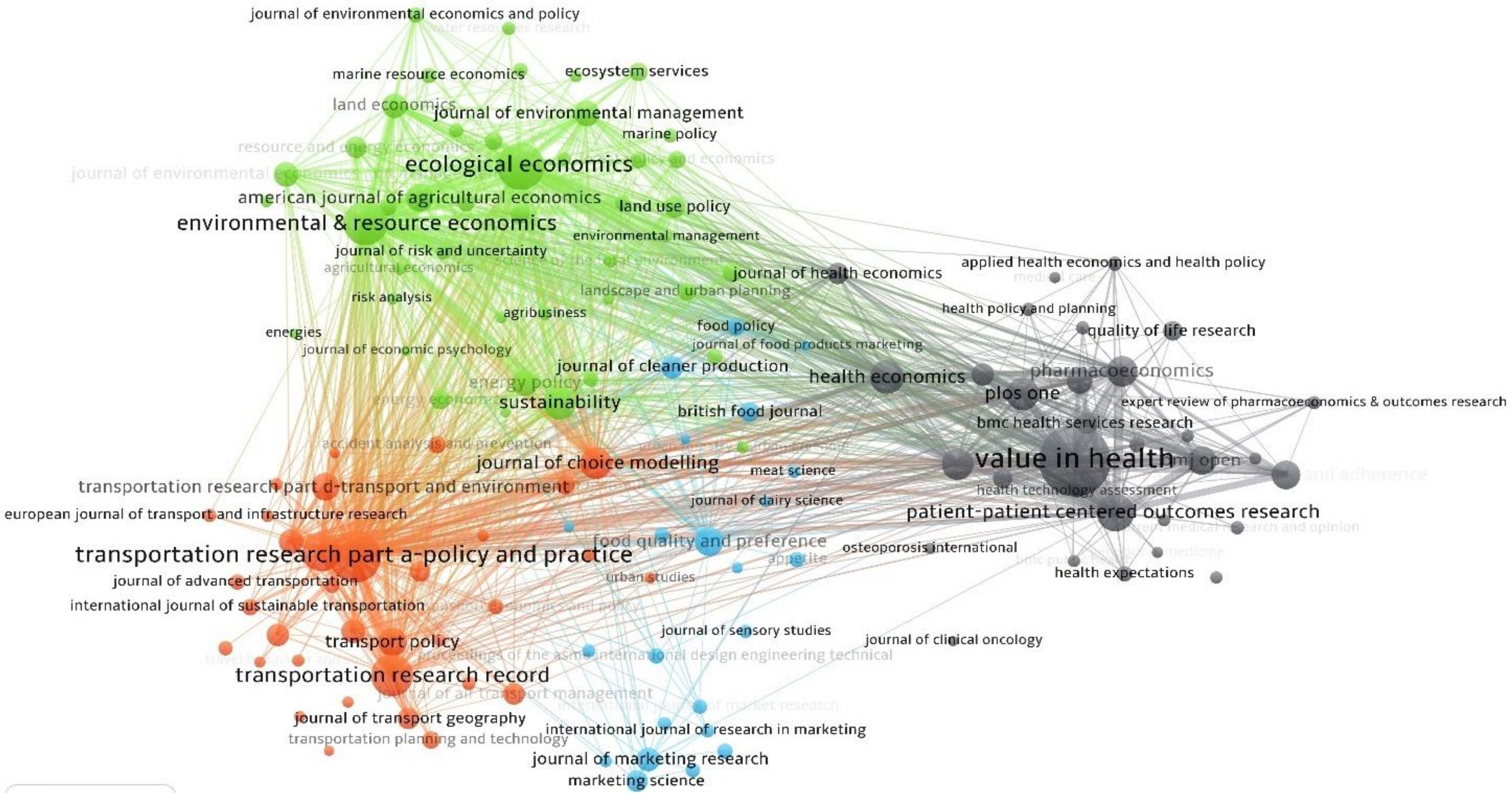

**Figure 3** Relationships of journals in the CE literature based on the similarity of the references of their publications.

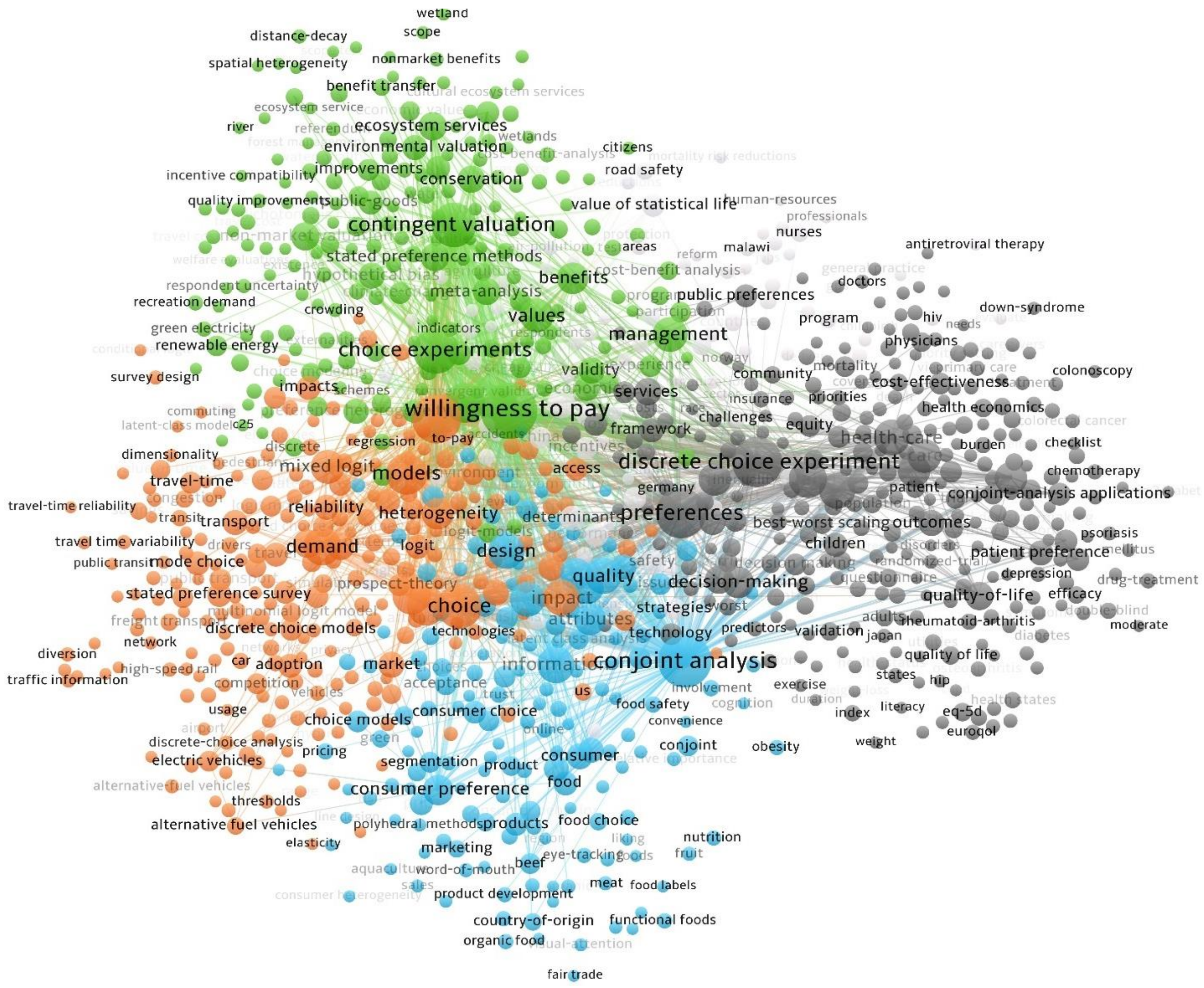

**Figure 4** The network of keyword co-occurance in the CE literature.

The term "*hypothetical bias*" appeared in the keywords of n=141 CE publications. Its highest frequency of co-occurrence is with the terms "*stated preference*", "*valuation*", "*choice experiments*", "*willingness to pay*", "*cheap talk*", "*values*", "*meta-analysis*", "*public goods*", "*contingent valuation*" and "*mechanisms*". "*Hypothetical bias*", as a keyword, has co-occurred 13 times with the keyword "*validity*". The keyword "*validity*" itself, in addition to appearing in its pure form, has also appeared in three other forms: "*validation*" (n=41), "*external validity*" (n=25), and "*convergent validity*" (n=18). While according to Figure 4, the keyword "*hypothetical bias*" most closely identifies with environmental valuation studies, it is clear that the use of the term is not exclusive to that domain.

To investigate the use of terms within each discipline, separate keyword analyses were undertaken for each of the disciplines, resulting in the density view maps in Figure 5. They reveal that environmental valuation studies have used the terms "*hypothetical bias*" and "*validity*" more frequently than any other domain of applied economics. The term "validity" appeared most often as a keyword in health and environmental studies, whereas consumer studies were the second most frequent users of the term "*hypothetical bias*" in their keywords. Health economists have tended to describe the problem as "*validity*" rather than "*hypothetical bias*" (with a ratio of nearly 1 to 4). Six instances of co-occurrence of "*hypothetical bias*" and "*validity*" appeared in environmental studies, the highest among the four disciplines. CE-related studies in the transport domain have used the two terms markedly less frequently than the other domains. Therefore, there is likely scope for transport researchers to learn about HB mitigation strategies in other disciplines.

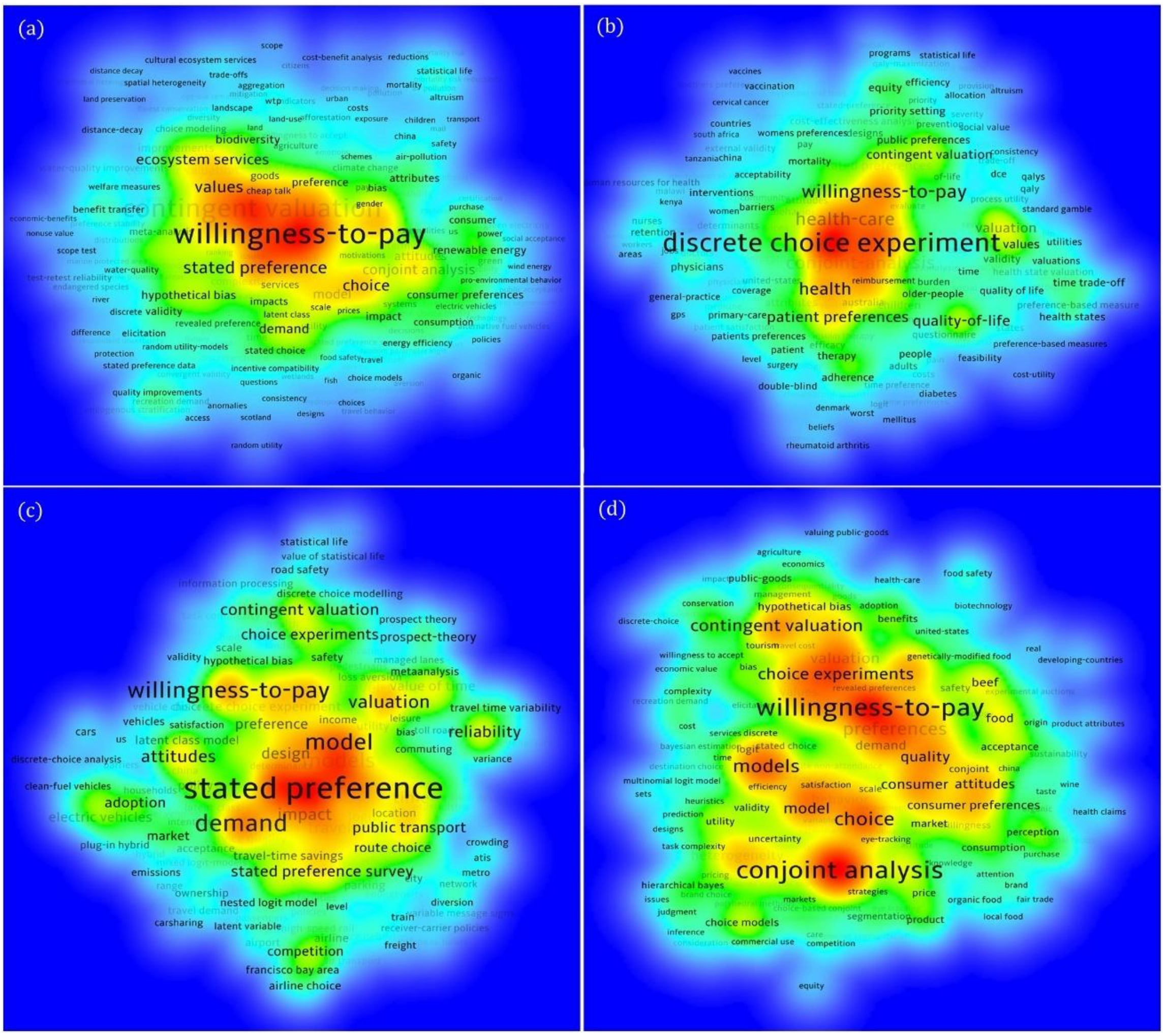

**Figure 5** Map of keyword co-occurance in (a) environmental, (b) health, (c) transport and (d) consumer studies of CE.

## 4. Mitigation strategies for hypothetical bias

Methods of mitigating HB generally differ based on whether they are applied during the design and administration stages of the survey in order to counter inherent sources of bias, or through follow-up questions whose information can be used to correct HB in model estimation. The former are *ex-ante* and the latter are *ex-post* measures of bias mitigation (Hofstetter et al., 2020; Whitehead and Cherry, 2007). The experimental literature that has assessed the effectiveness of these methods has employed two general approaches. The first is the theoretically ideal method and forms an evaluation of the level of *absolute hypothetical bias*. It consists of conducting two CEs with two groups of respondents, applying the mitigation method to one group and treating the other group as the control group, that is, administering the survey in the absence of a bias mitigation measure. The effectiveness of the bias mitigation method can then be evaluated objectively on an absolute basis, provided a bias-free (or less-bias-prone) set of observations/estimates (i.e., those of a more realistic nature) is accessible as the benchmark of comparison. The second approach evaluates the level of *relative hypothetical bias*, in which the evaluation is undertaken in the absence of any bias-free benchmark. Rather, assuming that the direction of bias is known *a priori* theoretically or based on previous experimental evidence, one can directly compare estimates, often of willingness-to-pay (WTP), across mitigation and control groups and interpret any differences as a measure of effectiveness of the bias mitigation method.

The empirical literature on bias mitigation methods shows a mixture of reliance on absolute and relative HB measurements, with a nearly equal split. When estimates with and without mitigation method are significantly different, then inferences about the effectiveness can be made. Such observation can be regarded as evidence for the existence of bias, and the observed differences can be interpreted as the extent of effectiveness of the mitigation method. However, it is not clear whether a failure to observe significant differences between mitigation and control groups should be attributed to the ineffectiveness of the mitigation method in reducing an existing bias or the absence of HB in the first place. This is a fundamental downside of testing the effectiveness of mitigation methods based on relative measures of HB.

It should also be noted that in absolute evaluations of mitigation methods, similar to the empirical tests of HB itself, the assumption of bias-free data cannot be made without compromise. More often than not, the benchmark that is treated as the source for *real* (bias-free) estimates is itself experimental and subject to various potential forms of bias even though they may be more realistic than purely hypothetical surveys. Examples are simulated markets/systems, field experiments, binding or incentive-aligned experiments, and self-reported revealed preference (RP) surveys. Using naturalistic data as the benchmark in studies of HB in CEs is rare (Haghani et al., 2020). Furthermore, it should be noted that methods of bias mitigation are not mutually exclusive and are often not studied in isolation. Ex-post and ex-ante methods can be employed in combination as complements (Whitehead and Cherry, 2007). Also, multiple ex-ante methods could be implemented concurrently, as many empirical CE studies have done. In such cases, unless the design allows so, one cannot disentangle the effect of a single mitigation method from the others, and the outcome of the study provides only an indication of their combined effectiveness.

Our survey of the literature identified 56 empirical investigations and ten different categories of HB mitigation methods, namely:

1. cheap talk;
2. choice certainty scales;
3. honesty priming;
4. induced truth telling and inferred valuation;
5. solemn oath;
6. opt-out option or budget reminders;
7. time-to-think method;
8. RP-assisted estimations;
9. referencing and pivot (contextually realistic) designs; and
10. perceived consequentially scales or consequentiality scripts.

The majority of these empirical investigations have been conducted in the contexts of consumer choice (n=21) and environmental/resource valuation (n=14), whereas experiments of HB mitigation in transport (n=10) and health (n=8) domains were less frequent. The methods are reviewed in the following sections. Appendix A provides a synthesis of the main components of the 56 studies included in the core analysis.

### *4.1. Cheap talk*

The *cheap talk* method was originally proposed by Cummings and Taylor (1999) to counter HB in CV surveys. It was characterised as a way to "directly induce subjects to provide responses to hypothetical valuation questions that correspond with responses observed when actual cash payments are involved" (p. 649). According to Cummings and Taylor (1999), the term was borrowed from the literature on bargaining and game theory, where it had been used in reference to "nonbinding communication of actions by two or more players in an experiment prior to their hypothetical commitment" (p. 650). When used in the context of a SP valuation study, it can be regarded as a nonbinding communication between a researcher and survey respondents prior to the administration of the survey (Lusk, 2003)). The method seems to have been inspired by and derived from experimental studies of Neill (1995) and Loomis et al. (1994) who investigated the effect of including budget constraints in CV of public environmental goods. Rather than attempting to remove the bias by citing budget constraints, a cheap talk script relies on an explicit discussion of the bias and its implications to make respondents aware of its existence. The cheap talk method has generally proven successful in CV studies although evidence is not unanimous and its effectiveness has been shown to depend on factors such as the length of the script (Aadland and Caplan, 2003; Aadland and Caplan, 2006), the description of (the direction of) the bias (Aadland and Caplan, 2006), subject characteristics (List, 2001), payment level (Bateman et al., 2009) and payment vehicle (Brown et al., 2003).

Proven an often successful bias mitigation method in CV surveys of public and private goods (Ami et al., 2011), particularly for consumers that are not very knowledgeable about the good (Lusk, 2003), cheap talk was subsequently adopted for CE applications (Ladenburg et al., 2011) and is thus far the most commonly used bias mitigation method in choice elicitation. A cheap talk script includes three general components: (i) it describes, prior to administration of the survey, the HB phenomenon to participants, (ii) it introduces respondents to possible explanations of the bias, and (iii) it pleads to subjects that they respond to the upcoming hypothetical questions with the knowledge of such potential bias while treating the hypothetical scenarios as if they encounter them in real life.

A pioneer empirical test of cheap talk effectiveness in CEs is reported in Carlsson et al. (2005) who showed estimates of marginal WTP for food products to be significantly lower when cheap talk scripts had been applied. Several follow-up studies investigated effectiveness of cheap talk scrips in relation to consumer choices of private goods (often food products). The results overall indicate successful mitigation of HB, in particular regarding experiments that measured relative bias. But evidence is mixed when one considers investigations of absolute bias. Utilising a large-scale online survey of apple product choices, Tonsor and Shupp (2011) observed that cheap talk scripts reduced estimated WTP and enhanced reliability of the estimates at the same time by narrowing the confidence intervals. They further suggested that a cheap talk script is more effective with respondents who are unfamiliar with the attributes. This different effect of cheap talk for various cohorts of consumers is in line with the findings of Lusk (2003) in relation to CV surveys. In the context of food choice with health implications, Chowdhury et al. (2011) observed that a cheap talk script successfully reduced absolute HB but did not fully eliminate it. Silva et al. (2012), using an incentive-aligned experimental setting, investigated the role of perceived task complexity on the effectiveness of cheap talk and concluded that the script is only effective when participants consider the task as easy. In another investigation of absolute HB mitigation using a field experiment as the benchmark, Moser et al. (2013) found that, although a cheap talk script reduced HB in WTP for most attributes of apple products, these reductions were statistically insignificant for most attributes. In a medical treatment choice context, Özdemir et al. (2009) found that a cheap talk script reduced WTP for almost all attributes, but to varying degrees. In the context of non-market public goods and services, List et al. (2006) demonstrated the effectiveness of cheap talk while also providing evidence that it may also induce inconsistency in subjects' preferences.

### *4.2. Choice certainty scales*

*Certainty scale calibration* can be regarded as the most-established *ex-post* correction method for reducing HB in CEs. Like most other mitigation methods, it originated from the CV domain. This approach is based on follow-up questions that measure respondent certainty about their choices or stated WTP values on a numerical scale (Champ and Bishop, 2001; Champ et al., 1997; Ethier et al., 2000; Johannesson et al., 1999; Poe et al., 2002) or categorical scale (e.g., "fairly sure", "absolutely sure) (Blumenschein et al., 1998; Blumenschein et al., 2001). For example, Champ et al. (1997) noted that while hypothetical donations significantly exceeded real donations, there was no significant difference (bias) when subjects were very certain of their yes responses. Similar promising findings were reported for other CV methods (Blumenschein et al., 2007; Harrison and Rutström, 2008), including referendums (Morrison and Brown, 2009), dichotomous choice experiments (Brouwer, 2011; Vossler et al., 2003) and auctions (Furno et al., 2019).

Adopting choice certainty calibration in the domain of choice experiments for environmental valuation, Lundhede et al. (2009) presented different ways of handling uncertain responses, including accommodating stated uncertainties in the scale parameter and treating it as a function of response uncertainty. In a second approach, they treated the scale parameter as a function of the specific variables found to influence stated uncertainty. Comparing against a benchmark model that discards stated uncertainties, they observed that the certainty scale correction results in more reliable estimates although the influence on WTP estimates per se was found to be insignificant. In a valuation of flood risk reduction, Dekker et al. (2016) also observed a correlation between choice uncertainty and randomness, with uncertain respondents making more random choices. Uncertain respondents were also observed to select the opt-out option more often than certain respondents. But contrary to the assumed purpose of choice certainty calibration, WTP for flood risk reduction increased after accounting for decision uncertainty. Relying on the well-established assumption of upward HB in public good valuation, this observation translates to magnifying HB as opposed to mitigating it. In another environmental valuation context where HB in WTP estimates was observed to be significant, Ready et al. (2010) suggests that they were able to largely mitigate this bias through respondent certainty calibration. In their approach, uncertainties are reflected in an additive component superimposed on the conventional error structure of the utilities.

In relation to certainty indexing, Beck et al. (2013) similarly show that a portion of idiosyncratic errors in choices can be explained by stated certainties, but also argue that model fit improvements or increases in the reliability of estimates (reduced standard errors) (Hindsley et al., 2020; Kunwar et al., 2020) may not necessarily reflect better behavioural representation and that econometric differences should, in that sense, be interpreted with caution (particularly in the absence of benchmark estimates as validation reference points). Similar concerns have also been voiced by Bobinac (2019) who suggested that “post-estimation uncertainty scores are malleable” (p. 75) and can be significantly correlated with entirely irrelevant information. As noted by Beck et al. (2013), there is currently no consensus on how certainty calibration should be applied to model estimation (Ku and Wu, 2018; Kunwar et al., 2020). Beck et al. (2016) examined three methods proposed in the literature for calibrating choice experiments via (i) reported choice certainty (recoding uncertain responses into the status quo, (ii) a weighting approach (Beck et al., 2013), and (iii) joint “choice and certainty” estimation approach (Rose et al., 2015) and concluded that incorrect calibration methods could even aggravate HB as opposed to mitigating it. Regier et al. (2019) also argue that variability of choice certainty is an important factor to be considered. They present a framework for identifying deliberative respondents by combining respondents’ certainty with their variability in certainty across a set of choice tasks. Recent studies have also investigated possibilities of inferring respondent’s objective uncertainty using measures such as eye tracking and response time as opposed to their subjectively stated degrees of uncertainty (Uggeldahl et al., 2016).

### *4.3. Honesty priming*

*Honesty priming* engages respondents in simple tasks that implicitly (covertly) primes them for honesty, i.e., it subtly and automatically activates their sense of honesty as opposed to explicitly asking them to give truthful answers. The term was borrowed from the social psychology literature (Pashler et al., 2013; Rasinski et al., 2005) and has been applied to various preference elicitation methods (De-Magistris et al., 2013; Gschwandtner and Burton, 2020; Howard et al., 2017; Liebe et al., 2019). The method relies on social psychology studies suggesting “priming” (i.e., incidental exposure to words or cues unrelated to the task) can unconsciously influence perception, behaviour and decision-making of people (Banerjee et al., 2010; Chartrand et al., 2008). A wealth of experimental findings in social psychology suggests that people can be primed to display behaviours such as fairness (in price negotiations) (Maxwell et al., 1999) or cooperation (in social dilemma games) (Drouvelis et al., 2010). Rasinski et al. (2005) experimentally demonstrated in that requiring people to complete a vocabulary task involving four words related to honesty (“honest”, “open”, “sincere” and “truthful”), embedded among other words, made them more likely (compared to those exposed to neutral words) to later admit having engaged in socially undesirable behaviour (excessive alcohol consumption). Further studies have produced evidence for various versions of priming honesty. It has been shown, for example, that primed with religious representations (religious words), experimental subjects cheated significantly less on a subsequent task (Randolph-Seng and Nielsen, 2007).

A pioneering application of honesty priming in CEs is reported in the study of De-Magistris et al. (2013). They showed, in a context of food product choice, that priming respondents for honesty, through a sentence scrambling test that preceded the CE, can significantly reduce marginal WTP estimates. They also observed that the marginal WTP of the hypothetical experiment with priming was comparable to that of an equivalent non-hypothetical treatment, which evidenced elimination of bias through the priming treatment. Their investigation also suggested that a cheap talk script alone does not completely eliminate HB, at least not compared to the priming effect. In choices of organic food products, Bello and Abdulai (2016a) observed that honesty priming and cheap talk both had significant impact on attribute non-attendance and that the marginal WTP estimates were significantly lower under honesty priming compared to cheap talk (while both being lower than the baseline with no mitigation measure). Further investigation by Bello and Abdulai (2016b) also produced evidence that honesty priming has a positive effect on survey engagement.

The existing empirical evidence does not entirely support the effectiveness of honesty priming in mitigating HB. For example, Gschwandtner and Burton (2020) reported for a food product choice context cheap talk to be much more successful in WTP reduction than honesty priming. In the context of environmental preservation policy choice, Howard et al. (2017) reported that online implementation of an honesty priming intervention resulted in no significant change in price sensitivity compared to that of the control group. They neither observed any significant effect of honesty priming in a face-to-face setting. Instead they observed that the effect particularly diminishes as the respondent proceeds further into the choice tasks. They found the cheap talk effect to have a larger effect on price sensitivity compared to honesty priming.

### *4.4. Induced truth telling and indirect questioning (inferred valuation)*

A semi-covert approach adopted for eliciting truthful preferences is *induced truth telling* which is a method founded in the *Bayesian Truth Serum* (BTS) method of Prelec (2004). Induced truth telling is a relatively new method for improving honesty and information quality in multiple-choice surveys and has been adopted by non-choice surveys too (Zhou et al., 2017). The difference with other ex-ante methods such as cheap talk and solemn oath is that it involves only an implicit request for truthful revelation of preferences. It is however a less subtle and more overt method of truthful preference elicitation than honesty priming.

BTS is a quantitative method for incentivising truthfulness to subjective survey questions. It rests on the assumption that subjects use their own opinions as signals about the distribution of opinions/preferences in the population (Barrage and Lee, 2010; Frank et al., 2017). The method utilises a deception-free information scoring system that induces truthful answers. It does not rely on a priori known distribution of responses and disincentivises responses towards the group mean. In this method, each respondent provides a personal answer as well as a prediction of the fraction of people endorsing that answer. Predictions are scored for accuracy and personal answers are scored by assigning high scores to answers that are more common than collectively predicted. Prelec (2004) showed that this makes truthful responding the only correct strategy even by those who are confident that their answers are a minority, as “one’s true opinion is also the opinion that has the best chance of being surprisingly uncommon” (p. 462) and that the truth telling is the Bayesian Nash Equilibrium (Prelec, 2004). The method does not require that the experimenter explains to respondents the mathematics underlying the scoring system, instead one would tell participants that answering truthfully will maximise their scores (and hence, their earned participation reward). A typical script can be found in the study of Barrage and Lee (2010) who were pioneers in adopting this method in CV surveys, while contrasting it with cheap talk and consequentialism. They observed that while real and consequentialism responses were statistically indistinguishable, the cheap talk and induced truth telling methods only eliminated bias for one of the tested goods. They also observed that the effect of BTS was more significant for females and more experienced subjects. Another application of the truth serum in CV studies has reported even more promising evidence suggesting that it outperforms more overt forms of truthful preference elicitation like the solemn oath, see Weaver and Prelec (2013).

There has been only a very limited number of applications of the induced truth telling and indirect questioning methods in CEs. A pioneering application of a closely related methodology in CEs consists of the studies of Lusk and Norwood (2009a) and Carlsson et al. (2010), who used a *third-party (indirect) approach* also known as the *inferred valuation* method (Lusk and Norwood, 2009b) to reduce HB, though neither makes any explicit reference to Prelec's BTS method. While this method is not technically the same as BTS, it presents strong parallels with it, in that they both share the common feature of eliciting people's predictions about others' valuations. Lusk and Norwood (2009a) reported the application of the inferred valuation method to consumer choices for goods with normative consequences (e.g., organic beef, environmental-friendly dishwashing liquid). They observed that participants indicated a higher WTP for themselves than for others. They concluded that when goods embody relatively high normative motivations, inferred valuation has the potential to lower HB. In Carlsson et al. (2010), subjects were asked to state how they believed the average respondent would answer choice questions regarding environmental donations. This treatment was compared with a more conventional cheap talk script, applied to own preferences. Marginal WTP estimates inferred from stated third-party preferences were observed to be significantly lower than those of own preferences. A within-subject investigation by Olynk et al. (2010) on consumer preferences for attributes of milk and pork meat revealed mixed evidence with respect to the effect of indirect questioning. Instead a comparable between-subject design resulted in indirect WTP estimates as small as 60 percent of those inferred from direct questioning (Klaiman et al., 2016). A recent study by Menapace and Raffaelli (2020) compared hypothetical choices with actual purchases at a grocery store and can be regarded as one of the first tests of the BTS in CEs, following the study of Dimitrov (2017). The study found that HB in consumer choices can be reduced, but not fully eliminated, using either third-party inferences or the BTS method.

### *4.5. Solemn oath*

Applications of solemn oath scripts, as an explicit mechanism for eliciting honest and truthful preferences, rest on the *theory of commitment* from social psychology (Charles, 1971; Joule et al., 2007; Kulik and Carlino, 1987; Wang and Katzev, 1990), which suggests that when a participant makes a promise in a hypothetical situation, they will be more likely to give an accurate biased-free answer. In this method, the investigator asks participants to swear on their honour to give honest answers to forthcoming questions. Stevens et al. (2013) have interpreted the solemn oath method "as an implicit contract between the researcher and the respondent" (p. 136). Applications have been reported in open-ended CV surveys (Stevens et al., 2013), dichotomous choice or referendum questions (Jacquemet et al., 2017), auction bidding questions (Jacquemet et al., 2013) and CEs (Carlsson et al., 2017; de-Magistris and Pascucci, 2014; Kemper et al., 2020; Lin et al., 2017; Mamkhezri et al., 2020).

The method is relatively new in SP applications. Jacquemet et al. (2013) were the first to report applications of oath scripts as a method of removing HB and found that taking the oath made people bid more sincerely in a hypothetical second-price auction, even more so than they did under monetary incentives. Subsequent CV studies have provided further evidence for the effectiveness of this method by showing that, under oath, the mean hypothetical and actual payments become indistinguishable (Stevens et al., 2013) and that people who sign an oath can become significantly less likely to vote for the public good in a hypothetical referendum (Jacquemet et al., 2017).

The existing investigations of this method in CEs are limited and do not provide clear and indisputable evidence as to whether oath scripts are an effective instrument for bias mitigation. Among the four studies that have, thus far, investigated this question, three have found WTP estimates to remain persistently unchanged despite oath administration (Carlsson et al., 2017; Lin et al., 2017; Mamkhezri et al., 2020). In a travel behaviour survey focused on travel time, comfort and cost, Carlsson et al. (2017) found no statistically significant difference between marginal WTP estimates associated with any of the attributes across the two subsamples, with and without oath scripts. Also, studying consumer preferences for solar energy panels, Mamkhezri et al. (2020) found no evidence that solemn oath lowers respondents' WTP. However, since these studies measure relative (potential) HB, it cannot be ascertained whether this failure to effect changes in WTP estimates is the result of an inherent ineffectiveness of oath scripts or due to the inexistence of HB in the first place. In other words, in the absence of true benchmarks for measuring absolute HB, these findings cannot be interpreted as unequivocal evidence for the ineffectiveness of oath scripts. For example, the study of Lin et al.

(2017) compared the effect of solemn oath with cheap talk, honesty priming and a control group, and found no significant differences between the different treatments. Among the existing empirical investigations of oath scripts in CE applications, the study of de-Magistris and Pascucci (2014) in the only one that observed significant lowering of WTP estimates. Their experiment was conducted in the context of insect-based food alternatives.

A related approach is based on *Query Theory* (Johnson et al., 2007), which suggests that preferences are constructed in the moment through a series of successive thoughts, rather than being pre-stored and instantly retrievable, making the final decision depend on the order of the thoughts. Testing this hypothesis, Kemper et al. (2020) assessed the differences in thought processes of individuals under oath and observed that an honesty oath changes the content and order of the queries.

### *4.6. Opt-out and budget reminder*

An important aspect of CE design concerns whether to include a form of "none" option in the choice sets. This option can be part of *forced* or *unforced* choice (Penn et al., 2019). When respondents are not given the opportunity to choose none of the options included in the choice set, the choices are regarded as forced (Penn et al., 2019). The unforced choice design could take the form of an *opt-out* or *status-quo* format (Kontoleon and Yabe, 2003). Evidence suggests that unforced designs, particularly in contexts that entail monetary trade-offs and consideration of budget constraints, could substantially affect WTP estimates (Penn et al., 2019).

Recognising the inherent differences between CEs and other CV methods, and arguing against the sufficiency of the simple adoption of cheap talk scripts for CEs, Ladenburg and Olsen (2014) suggested that we augment cheap talk scripts with an *opt-out reminder*. This method explicitly reminds respondents that they can choose the opt-out alternative, for example, if they find the experimentally designed alternatives too expensive. Such a reminder could be administered once; i.e. a single opt-out reminder (Varela et al., 2014), or at every choice set; i.e., a repeated opt-out reminder (Alemu and Olsen, 2018). The latter is meant to compensate for the diminishing effect of the cheap talk (or the initial opt-out reminder) as the respondent proceeds through the sequence of choice sets. In an experimental investigation, Ladenburg and Olsen (2014) found that implementing a repeated opt-out reminder had a significant impact on the total WTP estimate for the provision of a public good (i.e., significantly affecting the opt-in rate), while the effect on marginal rates of substitution was not substantial. In an experimental testing that allowed for the measurement of absolute HB by including a non-hypothetical setting, Alemu and Olsen (2018) demonstrated that a repeated opt-out reminder completely eliminated or mitigated the bias in marginal valuations for various attributes of novel food products. The investigation of Varela et al. (2014), however, suggested that a (single) opt-out reminder did not affect participation rates in forest preservation programmes beyond the effect of a cheap talk script, creating a mixture of evidence on this bias mitigation method. In relation to the latter finding, however, it should be noted that while Varela et al. (2014) compared the effect of cheap talk as well as "cheap talk + opt-out reminder" with that of a baseline (i.e., no mitigation method), they did not include a separate "opt-out reminder" treatment (independent of the effect of cheap talk). Therefore, it is not evident from their findings whether the neutral effect of the opt-out reminder arises from the inherent ineffectiveness of this method in influencing respondent's choices or is due to the fact that the cheap talk had already removed the potential HB. In other words, while their finding may suggest that a single opt-out reminder does not have any effect on further reinforcing a cheap talk script, it does not make it clear whether it could be used as a possible substitute for cheap talk, or whether a repeated version of the reminder could have had a more tangible impact.

In line with the notion of augmenting cheap talk scripts with an opt-out reminder, Gschwandtner and Burton (2020) proposed inclusion of a *budget reminder*. They compared the effect of a budget constraint reminder combined with a cheap talk script with that of an honesty priming treatment in the context of organic food choices. While both methods, to varying degrees, reduced HB, the explicit script (i.e., cheap talk + budget reminder) appeared to be more effective than the covert method, i.e., honesty priming. To our knowledge, the independent effect of a budget reminder or its effect relative to that of a cheap talk or to that of opt-out reminders have not, so far, been investigated in CEs.

*4.7. Time-to-think method*

It has been suggested, predominantly by health economists, that giving people more time to reflect/deliberate on their responses in CEs could be regarded as an alternative way of mitigating HB. Like most other mitigation methods, earlier applications of this method can be found in the CV literature, (Whittington et al., 1992). In a pioneering application of this method in CEs, Cook et al. (2007) gave half of their respondents one night to think about vaccine choice options presented to them in the survey. Compared to the sub-sample that was not given the extra time, respondents under the time-to-think treatment showed lesser instances of violating internal validity criteria and also indicated lower WTP values in their choices, a sign that the method was effective in reducing HB. This observation on the effect of extended deliberation time (Rigby et al., 2020) has been to large degrees replicated by further follow-up empirical testings in health-related choices (Ozdemir, 2015) suggesting that the method could reduce WTP estimates by up to approximately 40 percent (Cook et al., 2012). A seeming exception to this stream of relatively congruent findings is the study of Tilley et al. (2016), which used a split-sample design to estimate *willingness to accept* (WTA) of participants for different cash transfer programs aimed at improving public hygiene in Africa. Notable differences were observed in respondents' stated choices when given time to think, including lower stated WTA compared to those who answered immediately. A subsequent comparison with actual take-up in the real world revealed that both sub-samples had, in fact, underestimated the actual WTA. Taking the observed take-up rate as the bias-free benchmark, this would indicate that the time-to-think subsample produced even more biased estimates of WTA.

*4.8. Pooled estimation with RP*

Combining hypothetical choice data and RP data to improve model fit and estimation accuracy is a common practice in choice modelling. Particularly in transport and travel behaviour studies this method and its various theoretical implications have been heavily discussed in a broad range of contexts and applications including commuter valuations of travel time savings and reliability, preferences for transport modes or route choice (Abildtrup et al., 2015; Ben-Akiva et al., 1994; Cherchi and Ortúzar, 2002, 2006; Duann and Shiaw, 2001; Fifer et al., 2011; Haghani and Sarvi, 2017, 2018, 2019; Helveston et al., 2018; Hensher et al., 2008; Lavasani et al., 2017; Morikawa, 1994; Polydoropoulou and Ben-Akiva, 2001; Train and Wilson, 2008; van Essen et al., 2020; Wardman, 1988). This approach, however, has not been conventionally regarded as a mitigation method for HB. In fact, a portion of studies that have investigated HB, have used the model estimated on RP data as the benchmark for evaluating the extent of bias (Hensher and Bradley, 1993), rather than combining the two sources as a way of bias mitigation. Some authors, however, have recommended that such practice, i.e., estimating models on a combined datasets when RP's are available, could per se be a way of reducing HB, as an alternative ex-post method. Herriges et al. (1999), for example, have suggested that "rather than treating stated preference (SP) and revealed preference (RP) as competing valuation techniques, analysts have started to view them as complementary, where the strengths of each approach can be used to provide more precise and possibly more accurate benefit estimates" (p. 6). Whitehead et al. (2008) similarly pointed out that "Combining SP data with RP data grounds hypothetical choices with real choice behaviour" (p. 877) and that "Combination with RP data can be used to detect and mitigate hypothetical bias and validate SP methods" (p. 877).

Previous research has demonstrated that the use of hypothetical choice data yields attribute valuations and marginal rates of substitution comparable to that of counterpart RP. But other important metrics such as total WTP and market share estimates (often embedded in the estimates of alternative-specific constants), elasticities and parameter scales could suffer more tangibly from the inherent disparities between RP data and data from CEs (Hensher et al., 1998; Hensher, 2008; Hensher and Li, 2010; Louviere et al., 1999; Resano-Ezcaray et al., 2010; Swait et al., 1994). A jointly estimated model could then be a remedy for these sources of bias (Hensher and Bradley, 1993).

The recent study of Buckell and Hess (2019) takes such perspective by linking the approach of supplementing CE with RP data directly to the issue of EV and HB and recommending this practice as a potential remedy for weak EV. In the authors' terms "Revealed preference (RP) data do not suffer from hypothetical bias. Thus, if available, incorporating RP data in choice models can abate hypothetical bias in model estimates and the derived metrics such as forecasts" (p. 94). They proposed methods for correcting scale as well as alternative-specific constants of utilities using RP in a health-related context (smoking habits) and observed that such corrections could make substantial differences to the forecasts. An earlier application in health studies can also be found in Mark and Swait (2004) where they demonstrated how the utility scale can be corrected in a joint estimate of physicians' preference for alcoholism medication. Evidence from similar studies that have used this method predominantly point out that a jointly estimated model often outperforms the hypothetical model and provides more accurate estimates, hence suffering less from the issue of HB. A major hinderance to the frequent use of this method in eliminating HB in CEs, despite its established usefulness and intuitive benefits, is its conditionality on the availability of RP data, which, a condition which more often than not is not met, in particular for non-market or novel goods.

### *4.9. Referencing or pivoting and (contextually) realistic design*

A major source of HB in CEs is the lack of contextual tangibility or the respondent's lack of familiarity/experience with the good or service in question (Schläpfer and Fischhoff, 2012). As a result, it is intuitively understandable that any measure taken towards grounding CE surveys more in reality, in terms of providing context, could be a way of countering the effect of HB. A method that has particularly been cited as a potential solution to the lack of contextual realism in choice surveys is *pivoting* or *referencing* to a real experience (Hensher et al., 2012; Li et al., 2018; Rose et al., 2008; Train and Wilson, 2008; Train and Wilson, 2009). In this design paradigm, attributes of the alternatives are constructed relative to a respondent's experienced/chosen alternative in the real world. This is in contrast with the more conventional methods of CE design where all choice sets for all participants are designed as a priori, irrespective of the individual respondent's experience in real world, i.e., a single fixed design for all. The notion of tailoring choice sets to real experiences and pivoting the design around a reference alternative has been regarded as one that "appears to offer promise in the derivation of estimates of WTP that have a meaningful link to real market activity, closing the gap between RP and SC WTP outputs" (Hensher, 2010) (p. 735). Unlike the pooled estimation approach, this approach is not subject to the availability of an RP dataset. Information about a chosen (experienced) alternative would be sufficient to generate a pivot design. For example, in the SP-off-RP paradigm proposed by Train and Wilson (2008), hypothetical choice sets are built by worsening the attributes of the chosen alternative and/or improving the attributes of the non-chosen ones. This exempts the analyst from constructing the non-chosen RP alternatives and their attributes, which is arguably the most challenging aspect with respect to the use of RP data in many contexts. Moreover, the referencing method can even be integrated with the principles of efficient survey design (Rose and Bliemer, 2009) to make use of the advantages of both contextual/behavioural realism and statistical efficiency at the same time. Rose et al. (2008) outline practical ways for such integration between pivot and efficient designs, such as the notion of constructing adaptive personalised choice sets (Fowkes and Shinghal, 2002). The only disadvantage could be that referencing may complicate the estimation procedure by introducing endogeneity issues to the choice process. However, proper estimation methods have been proposed to account for this issue (Danaf et al., 2020; Guevara and Hess, 2019; Train and Wilson, 2008; van Cranenburgh et al., 2014).

Despite the evident advantages of reducing HB and the slowly growing use of the referencing method in survey applications (Haghani et al., 2015; Hasnine et al., 2017; Hess, 2008; Hess and Rose, 2009; Masiero and Rose, 2013; Rose and Hess, 2009; Yu et al., 2013), empirical tests of the effectiveness of this method have been extremely limited. Recently, Chiu and Guevara (2019) put into test the SP-off-RP method of Train and Wilson (2008) in a commuter mode choice by comparing SP, SP-off-RP and RP designs, and investigated whether the SP-off-RP questions can reduce HB. They observed that the Value of Time estimates inferred from conventional SP responses underestimated the real values, while the outcomes associated with the SP-off-RP and RP data were statistically indistinguishable. Moreover, the SP-off-RP model demonstrated a better predictive ability on a hold-out RP sample.

In line with previous discussions on potential merits of pivoting the design around an experienced alternative (Bradley, 1988; Matthews et al., 2017), one may also identify other similar avenues that could potentially be utilised as a bias mitigation method through providing context and tangibility. It may be argued that reducing *task complexity* (Arentze et al., 2003; de Bekker-Grob et al., 2019; Hensher, 2006; Meyerhoff et al., 2015; Swait and Adamowicz, 2001) or enhancing presentation format (Rossetti and Hurtubia, 2020) could contribute to reducing HB and ultimately improving the EV of outcomes (de Bekker-Grob et al., 2020). Empirical testings regarding the potential effect of task complexity on the magnitude of HB are scarce. The study of Caussade et al. (2005) varied elements of task complexity (including the number of available alternatives, range of attribute levels and number of choice sets) systematically while measuring the effect on respective WTP estimates. Another relevant work is Meyerhoff and Liebe (2009), who showed that perceived task complexity can influence the choice of status-quo alternative (hence, potentially impacting total WTP estimates). Studies also considered the correlation between task complexity and error variance rather than task complexity and HB (Caussade et al., 2005; Lu et al., 2008). In an indirect test, Silva et al. (2012) show how perceived task complexity can diminish the effectiveness of cheap talk, suggesting at least indirect potential interplay between task complexity and the extent of HB. Furthermore, many of the works that have tested the effect of task complexity on WTP estimates cannot be characterised as tests of HB mitigation because of the ambiguity that often exists with respect to the direction of (potential) HB (Hensher, 2006).

Another dimension of behavioural realism pertains to how information (given a fixed level of task complexity) is presented to respondents. CE surveys are commonly administered using web-based formats in which choice scenarios are presented as a static screenshot that show the alternatives and the numerical values of their attributes. More recently, there has been a tendency to enhance the presentation of attributes using various forms of basic graphics (Oppewal et al., 1997) such as visualisation of distributions and percentages (de Bekker-Grob et al., 2020). A step forward in the direction of enhanced contextual realism that has been discussed by a number of recent studies is the application of virtual reality (VR) technology and 3D videos (Matthews et al., 2017; Rid et al., 2018; Romero et al., 2017), driving simulators (Fayyaz et al., 2020; Hess et al., 2020), maps and GIS information (Yamada and Thill, 2003) in choice task presentation. While VR experiments have tested various dimensions of data quality (such as measures of internal consistency and error variance (Bateman et al., 2009)), these experiments have not yet been treated as tests of HB mitigation. For example, Bateman et al. (2009) have demonstrated that the VR treatment significantly reduced the differences between WTP for gains and WTA for losses compared to a standard presentation format. Whether or not these technological developments could engender real differences in the ecological validity of CEs and whether they are warranted to be used as a way of countering HB remain open questions.

Another dimension to be considered is the potential that exists in choosing the most suitable *survey vehicle* in minimising HB. Knowing whether an interview survey is more successful than a web-based counterpart in eliciting deliberative and truthful responses could itself open further avenues for mitigating HB, see Sandorf et al. (2016). One, for example, could consider how administering a CE in the form of a face-to-face interview could potentially magnify the warm glow and social desirability effects (as discussed in Part I of our study) when public good valuation or goods with moral components are valued through the survey. These are all underexplored dimensions of choice survey presentation that could potentially play a role in bias reduction, hence warranting further empirical testing.

*4.10. Perceived consequentiality, real talk and consequentiality script*

The incentive to truthfully reveal preferences in a CE, also known as *incentive compatibility* (Buckell et al., 2020; Carson and Groves, 2007; Carson et al., 2014; Cummings et al., 1997; Zawojska and Czajkowski, 2017) or *incentive alignment* (Ding et al., 2005; Dong et al., 2010), is linked to whether subjects perceive their responses to be consequential (Vossler et al., 2012). According to Vossler and Watson (2013), a CE aimed at valuation of nonmarket public goods is deemed consequential if respondents care about the presented policy and view their responses as potentially influencing the decision regarding the implementation of the policy (also known as *policy consequentiality* (Herriges et al., 2010)). A natural implication of this assumption is that respondents who perceive the survey to be (policy-wise) consequential (to affect an outcome that they care about), will behave differently compared to those who believe their responses to the survey are of little to no policy consequence (Interis and Petrolia, 2014). This condition jointly with the condition of *payment consequentiality*, i.e. respondent perceiving that there is some probability that they need to pay depending on the choices that they make, are often referred to as *strong consequentiality* conditions (Herriges et al., 2010). According to Carson and Groves (2007), if a CE satisfies these conditions, that is, if a respondent believes that there is a positive probability that the survey is consequential in terms of both policy and payment, then their best strategy will be to answer truthfully (Herriges et al., 2010). This is often referred as the *knife-edge phenomenon* by economists, following the terminology of Carson and Groves (2007).

These considerations have opened two related avenues for bias mitigation. One is showing a *consequentiality script* whose purpose is to assure subjects of the real influence of their responses on policy implementation, constituting an ex-ante method. This has been shown in the CV literature of nonmarket valuation to have comparable effects to that of a cheap talk script (Bulte et al., 2005). A variation of this method has also been proposed in a private consumer good context, referred to as *real talk* scripts, where the participant is informed in a hypothetical valuation treatment that a non-hypothetical version of the study with similar but not necessarily identical goods will follow, prompting the subject to be consistent with their real preferences, as supported by the cognitive dissonance theory (Alfnes et al., 2010). The second approach would be measuring subjective self-evaluated degrees of *perceived consequentiality* through follow-up questions after the completion of the survey, an ex-post method that presents parallels to the certainty calibration approach (Herriges et al., 2010; Vossler and Watson, 2013).

In CEs, empirical evidence from both consumer studies (Lewis et al., 2016; Zawojska et al., 2019b) and public good valuations (Oehlmann and Meyerhoff, 2017) have suggested that highlighting the consequences of the choice survey, in the form of a script, increases the belief of participants in the potential policy consequences of their responses. This has been mostly established through subjective post-survey and self-reported measures of perceived consequentiality. But the evidence on how this impacts on the estimates is still rather mixed. Li et al. (2017) showed that increased belief in consequentiality of the choices (for provision of beef products) increased WTP. Lewis et al. (2016) showed that higher belief in consequentiality increased the likelihood of opting-in for purchase of genetically-modified labelled sugar. Oehlmann and Meyerhoff (2017) found no significant effect associated with highlighting consequences of the survey on estimated WTP for renewable energy systems. In a similar context, Zawojska et al. (2019a) differentiated between respondent perception of payment and policy consequentiality and observed that the two have opposite effects on price sensitivity and WTP estimates. While increased perception of policy consequentiality increased WTP for renewable energy, increased perception of payment consequentiality had the opposite effect. The suggestion of Lloyd-Smith et al. (2019) stating that “these [consequentiality] questions may not be a panacea for stated preference validity issues” is, in fact, an fitting reflection of the mixed and inconclusive evidence that currently exists on this class of HB mitigation methods.

## 5. The relation between sources of hypothetical bias and mitigation strategies

Our earlier investigation of empirical findings on HB in CEs (Haghani et al., 2020) suggested that slightly more than half of the studies that tested the HB problem have found clear signs of significant bias, while the other half is split between studies that have found insignificant bias or mixed evidence. In other words, evidence of significant HB has been found in nearly twice as many studies that have found negligible HB in CEs. Therefore, what we know so far is that although HB is not a universal problem across all contexts and applications of CE, the likelihood of its existence is undeniably high in CEs. This clearly highlights the significance of developing a nuanced knowledge base as an essential step for recognising survey applications in which the bias is more likely to exist. It also underscores the importance of implementing efficient strategies during the design of CEs to reduce the magnitude of potential HB. In doing so, understanding the chief sources of bias in each CE application that are most likely to cause deviation of stated choices from their true version would be key. In the previous work linked to this article (Haghani et al., 2020), we proposed a unifying conceptualisation of HB while also cataloguing and categorising an array of possible explanations or causes for HB that embodied different perspectives and existing theories on this topic in the economic and psychology literatures. Here, we argue that such recognition of HB sources could be critically instrumental in adopting effective bias mitigation strategies in CE design.

It is suggested that, in cases where HB is thought to mainly stem from respondents consciously or semi-consciously hiding of true preferences—such as inflating WTP to look socially desirable or portray a positive image of self or having a motivation to strategically distort the outcome or protest the survey—ex-ante measures that explicitly or implicitly encourage honesty could be critical in countering respondents' motivations for giving less-than-truthful answers. This includes, but is not limited to, ex-ante methods such as cheap talk, oaths or priming for honesty. In such circumstances, one could also assume that masking the purpose of the survey[1] (where possible, and particularly to counter deceit when provision/pricing of private goods are involved) or asking questions from the perspective of a third party (e.g., the inferred valuation approach) or incentivising the respondent with monetary reward for being as truthful as possible (e.g., the BTS method) could be other possible ways of keeping dishonest answers to a minimum.

In certain survey applications, particularly those concerning novel non-existent goods with which the respondent has no to little experience, the analyst should recognise that potential HB could be largely attributable to the lack of contextual tangibility or the fact that the respondent may not be able to effectively and accurately imagine the hypothesised context of choice or to predict the emotional states that they may experience, should the choice materialise. In such circumstances, the analyst may make efforts to provide enhanced contextual tangibility in the design and enrich the presented information or give the respondent extended time to ponder on their choices and subsequent feelings arisen from those choices. Alternatively, the analyst may decide to make a reference in the design to an already experienced alternative and/or incorporate certainty scales into the estimation procedure to reduce HB.

One could also envision many CEs where bias mainly arises from the fact that the respondent does not have to back-up their choices with actual payments or does not think the survey is of any financial/policy consequence and is only meant to serve an intellectual curiosity of the experimenter(s). In such cases, if incentive-alignment makes respondent's choice financially consequential, then the method could be adopted. For the purpose of this discussion however, we assume that an incentive-aligned design is not an option (which is the case in many survey applications). We consider cases, for example, where the stake is higher than what can be offered to respondents as house money, essentially prohibiting hypothetical choices to become financially binding unless one accepts the possibility of making the respondents financially worse off compared to when they started the survey. Or cases, where the good in question does exist in the market but is not in the possession of the experimenter to offer in the lab (e.g., purchase of electric vehicles). In such scenarios, and in the absence of RP data to assist the estimation, the most practical tools for the analyst would be resorting to cheap/real talk scripts or reminding the respondent of their budget constraints or the possibility of "not buying" any of the options, repeatedly throughout the survey.

[1] To our knowledge, we are not aware of any study that has previously tested masking of survey purpose as a potential way of reducing HB.

When there is enough evidence that a respondent may have strong scepticism about the societal impact of their responses to the survey, then presenting a consequentiality script that assures the participant of the significance of their responses for subsequent policy-making or incorporating perceived consequentiality measures in the estimation process could be pragmatic ways to counter the bias. Figure 6 summarises the above discussions and shows how various methods of HB mitigation may potentially counter a certain source of HB. While this could serve as a general guide, determination of the most effective method(s) for HB reduction needs be made on a case-by-case basis, and in recognition of the most likely sources of HB.

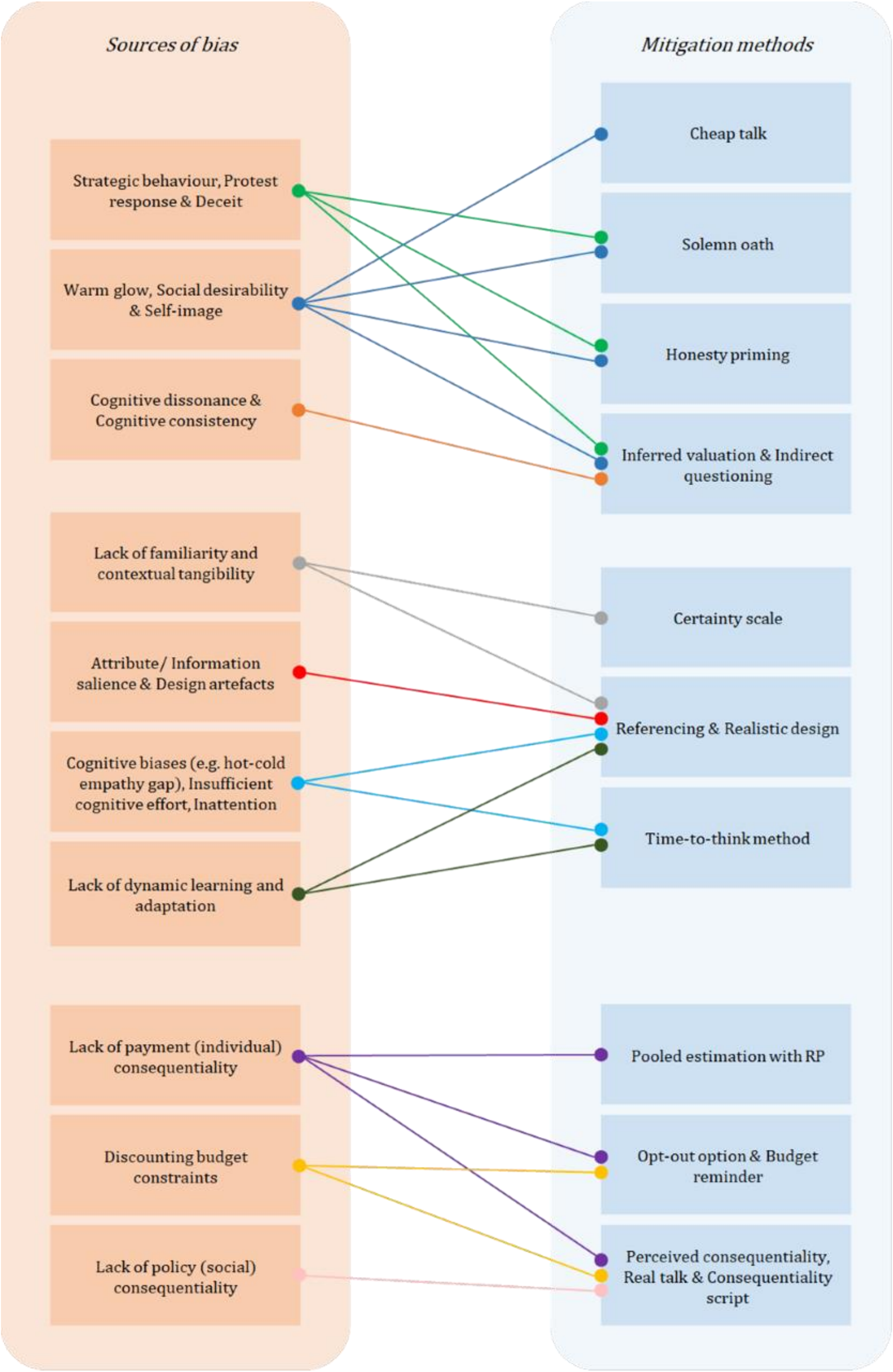

**Figure 6** Relations between causes/sources of hypothetical bias and mitigation strategies, i.e., which mitigation methods are best suited to counter each source of bias.

## 6. Prevalence of mitigation methods and studies

We reviewed the empirical evidence on HB and discussed a diverse array of empirically tested mitigation strategies for countering HB. Our review showed that different disciplines have focused on different subsets of these mitigation strategies. Figure 7 shows the break-down, for each sector of applied economics, of the number of studies that have tested each bias mitigation method. While empirical investigations of cheap talk effectiveness have been undertaken mostly by consumer and environmental economists, other methods each appear to have gained the attention of choice modellers within a certain discipline. For example, the ex-post methods of certainty scale and perceived consequentiality scales seem to have been most popular among environmental economists, whereas consumer economists have taken a notable interest in methods that encourage truthful responses, i.e., honesty priming, solemn oath and induced truth telling. Instead, the time-to-think approach was mainly used in health economics. Transport researchers have taken only a modest interest in realistic design methodologies as a way of countering the bias and their main way of doing so has been RP-assisted model estimations. Overall, while all four disciplines have engaged in empirical testing of HB, empirical investigations of mitigation strategies have received much less attention in transport and health than in consumer and environmental economics. Cheap talk has been the most popular and most frequently tested method of bias mitigation. In contrast, attention to how behaviourally realistic designs could be deployed as a way of countering HB has remained relatively limited (see Figure 8).

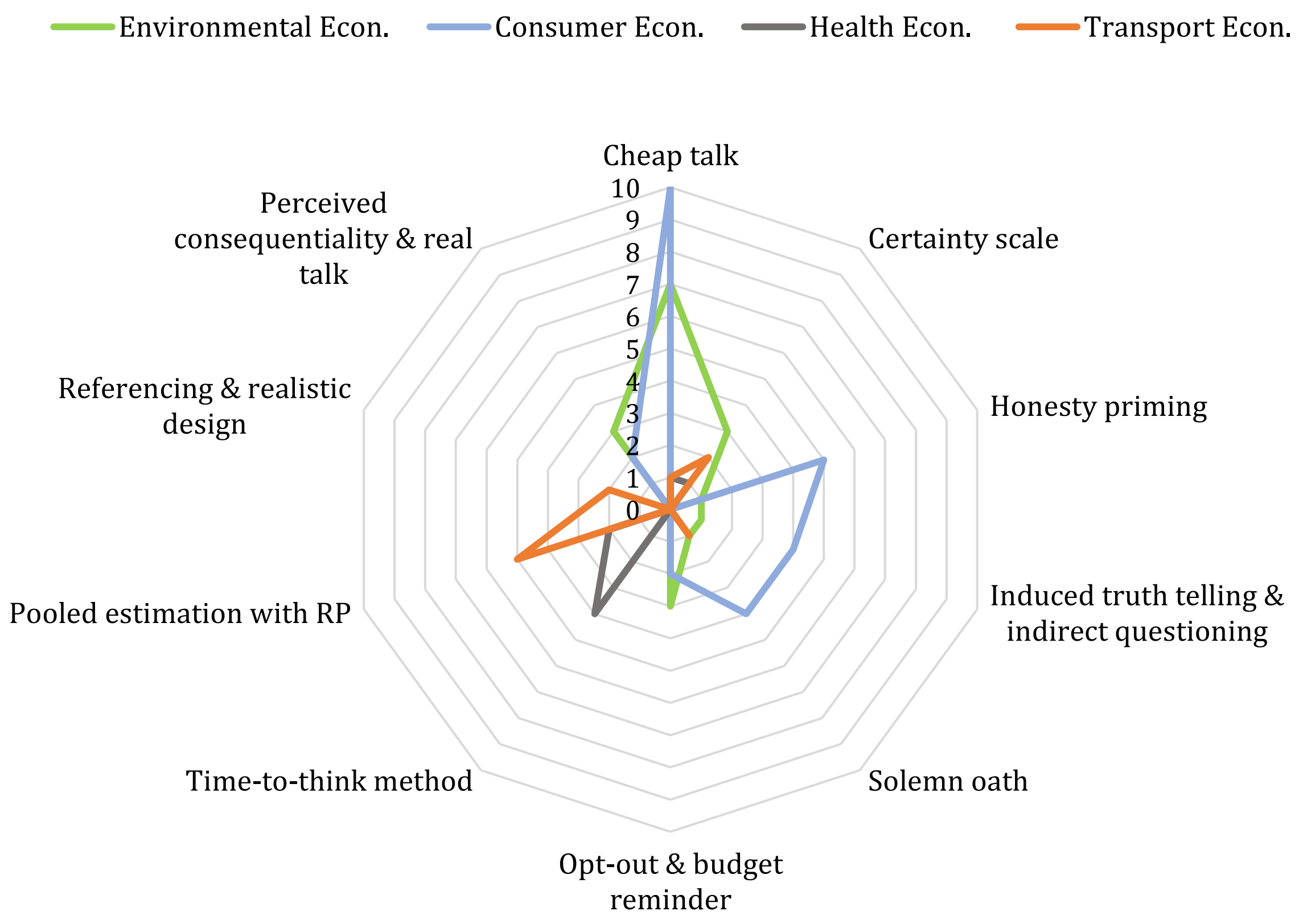

**Figure 7** Break-down of the frequency of empirical studies on the effectiveness of various hypothetical bias mitigation methods, by applied economics domain (environmental, consumer, health and transport).

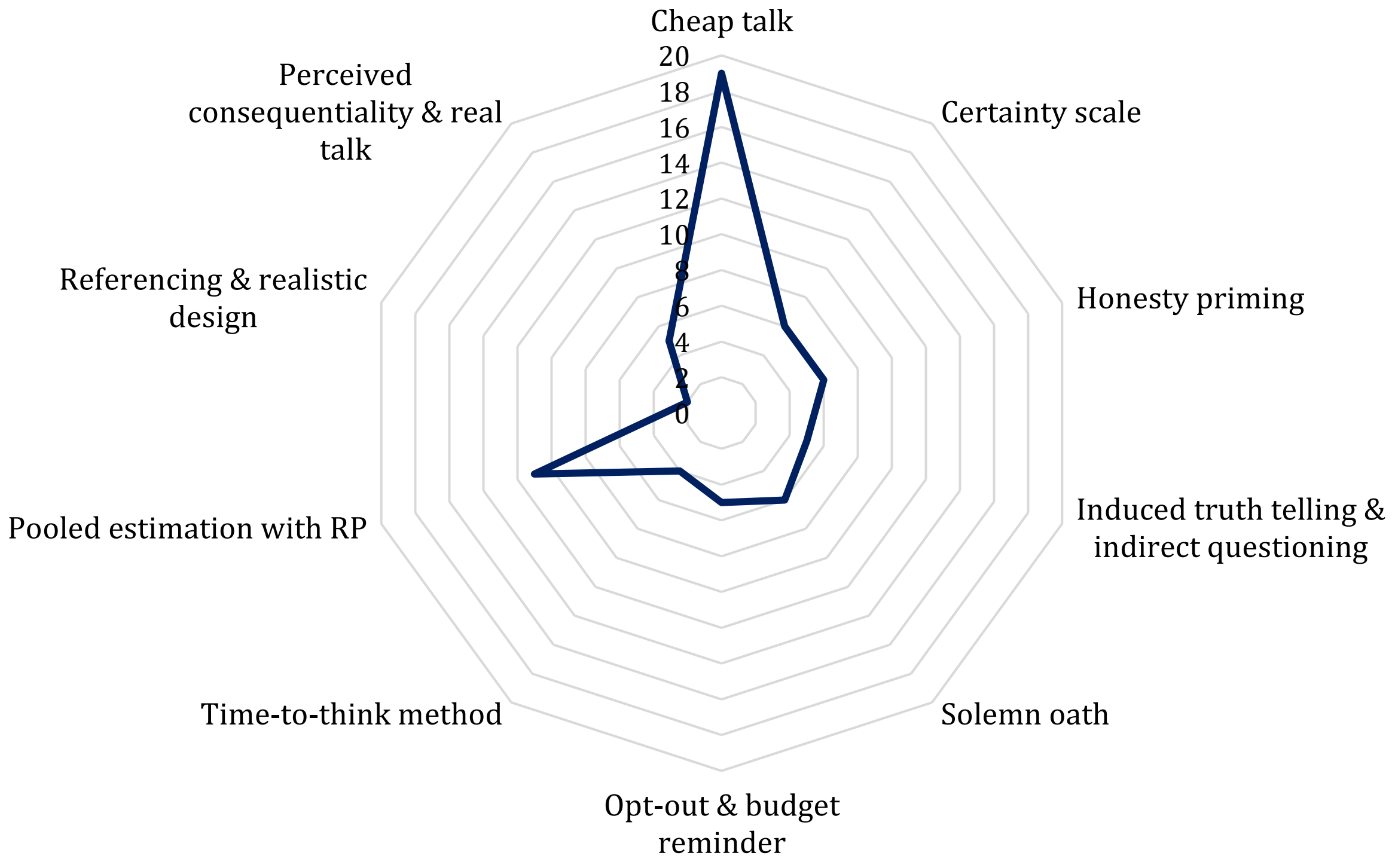

**Figure 8** The overall frequency of empirical studies on the effectiveness of various hypothetical bias mitigation methods.

## 7. Conclusion

The review in our earlier paper of the empirical evidence for Hypothetical Bias in Choice Experiments suggested that HB cannot be ignored. A central argument of the present paper has been that no single method has proven to be a panacea for mitigating HB. As Murphy et al. (2005) have pertinently pointed out, 'no single technique will be the "magic bullet" that eliminates this bias. Ultimately, mitigating hypothetical bias will probably involve a combination of techniques, including both instrument and statistical calibration' (p. 337). Having carefully reviewed the existing evidence on the effectiveness of these methods, we second that statement. We further suggest that these methods, although not invariably, to a large extent can be considered effective in reducing HB, especially when used in particular combinations where the methods complement each other. Our holistic overview of the existing literature also makes it clear that reducing HB does not necessarily require linking choices to financial payments. Lack of payment consequentiality is only one among several sources of HB, and when there is evidence of it being the main source of bias, then effective mitigation measures can be adopted, including making the choices financially consequential, if possible. But there are many applications where this method is irrelevant or impractical. The causes of HB are diverse and case-dependent, and so should be the ways to counter this problem. We suggest that the analyst recognises the sources of bias that are most likely at play in each application of CE, rather than simply accepting the possibility of bias and hoping for the best, and implements custom-chosen suitable mitigation strategies that best counter these sources of bias. Given the availability of a broad range of well tested and often readily applicable methods, it is recommended that the use of bias control strategies tailored to the particular structure and topic of the choice surveys should become a standard criterion for good practice in CE design.

**Acknowledgments**
This research was funded by Australian Research Council grants DP150103299 and DP180103718.

## Appendix A

**Table A1** Summary of studies on methods of mitigating hypothetical bias in discrete choice experiments

| Reference | Mitigation method | Choice context | Between/within Subject design | Absolute or Relative | Mitigation method effective? | Highlights |
|---|---|---|---|---|---|---|
| Carlsson et al. (2005) | Cheap talk | Food choice | Between | Relative | Yes | Estimated MWTP for food was lower in the survey version with cheap talk |
| List et al. (2006) | Cheap talk | Public non-market goods | Between | Absolute | Yes | - Responses were not statistically different between the real and hypothetical with cheap talk treatments<br>- Subjects in the hypothetical with cheap talk treatment were more likely to make inconsistent decisions |
| Özdemir et al. (2009) | Cheap talk | Medical treatments | Between | Relative | Yes | - Cheap talk not only affected the coefficient of the cost attribute, but also preferences for other attributes<br>- WTP estimates were generally lower in the cheap talk sample |
| Tonsor and Shupp (2011) | Cheap talk | Food demand (apples) | Between | Relative | Yes | - Cheap talk scripts not only influenced the level of WTP, but also may produced more reliable estimates<br>- The magnitude of the impact on WTP depended on respondent familiarity |
| Chowdhury et al. (2011) | Cheap talk | Food choice | Between | Absolute | Yes | - Results confirmed the presence of significant hypothetical bias<br>- Cheap talk reduced the magnitude of bias but did not fully eliminate it |
| Bosworth and Taylor (2012) | Cheap talk | Purchase a tree | Between | Absolute | Mixed | - A dramatically larger number of subjects opted-into the market in the hypothetical survey compared to the real payment condition<br>- Cheap talk induced respondents to opt-out of the market<br>- Participants in the hypothetical treatment with cheap talk were more price sensitive compared to the real payment treatment (cheap-talk overcorrection) |
| Silva et al. (2012) | Cheap talk | Food choice | Between | Absolute | Mixed | - Perceived task complexity had a significant impact on cheap talk's effectiveness in reducing HB<br>- The cheap talk script was effective only when subjects considered the task to be easy |
| Moser et al. (2013) | Cheap talk | Food choice (apples) | Between | Absolute | Mixed | - Results confirmed the presence of hypothetical bias<br>- Results confirmed the mixed effectiveness of a cheap talk script |
| Ready et al. (2010) | Certainty scale | Wildfire rehabilitation | Between | Absolute | Yes | - Hypothetical WTP was three times larger than the real estimate<br>- Certainty calibration successfully mitigated the bias |
| Broadbent (2014) | Cheap talk<br>+<br>Certainty scale | Recreation site expansion | Between | Absolute | No | - Hypothetical bias was not present in the MWTP valuation for the quasi-public good<br>- Cheap-talk and follow-up certainty were found to reduce MWTP estimates to be less than actual estimates |
| Fifer et al. (2014) | Cheap talk + Certainty scale | Driving behaviour | Between | Absolute | Yes | - SC model estimates were prone to HB<br>- Cheap talk and certainty scales when combined have the potential to compensate for HB |
| Beck et al. (2016) | Certainty scale | Driving behaviour | Between | Absolute | Mixed | - Incorrect calibration of responses can worsen the magnitude of HB<br>- By jointly estimating choice and choice certainty there is a significant reduction in HB |
| Dekker et al. (2016) | Certainty scale | Flood risk reduction | Between | Relative | No | - WTP estimate for a public good increased after accounting for stated choice uncertainty |
| Regier et al. (2019) | Certainty scale | Clinical treatment choices | Within | Relative | Yes | Respondents with higher mean and variability in certainty made choices that were more internally valid |
| De-Magistris et al. (2013) | Honesty priming + Cheap talk | Food products | Between | Absolute | Mixed | - MWTPs in the honesty priming treatment were significantly lower than those in baseline hypothetical CE<br>- Values from hypothetical CE with honesty priming were not significantly different from non-hypothetical CE<br>- Cheap talk script was not able to mitigate the HB in hypothetical CE |
| Bello and Abdulai (2016a) | Honesty priming + Cheap talk | Organic food product | Between | Relative | Yes | Honesty priming resulted in lower WTP values by nearly a factor of two relative to cheap talk for three of the four attributes |
| Bello and Abdulai (2016a) | Honesty priming + Cheap talk | Organic food product | Between | Relative | Yes | The level of survey engagement was higher under honesty priming effect compared cheap talk |
| Howard et al. (2017) | Honesty priming + Cheap talk | Environmental policy | Between | Relative | No | - The cheap talk effect faded with repeated choices<br>- Online implementation of an honesty priming intervention yielded no significant change in price sensitivity compared to a control |
| Lusk and Norwood (2009a) | Induced truth telling & indirect questioning | Consumer goods | Between | Relative | Yes | - WTP estimates for normative goods using inferred valuation could be twice smaller than that of the conventional valuation<br>- WTP estimate for goods with low normative motivations were similar across the methods |
| Carlsson et al. (2010) | Induced truth telling & indirect questioning + Cheap talk | Donation to environmental projects | Between | Absolute | Yes | - Both hypothetical treatments (own and third-person preference) showed large differences with the real-money treatment<br>- HB effect was smaller when using a third-person preference viewpoint |
| Olynk et al. (2010) | Induced truth telling & indirect questioning | Meat & dairy products | Within | Relative | Mixed | - Indirect questioning yielded statistically smaller WTP estimates for certain products and certain attributes |
| Klaiman et al. (2016) | Induced truth telling & indirect questioning | Consumer preference for packaging | Between | Relative | Yes | - Indirect questioning yielded significantly smaller WTP estimates |
| Menapace and Raffaelli (2020) | Induced truth telling & indirect questioning | Food choice | Between | Absolute | Mixed | - Inferred valuation and Bayesian Truth Serum both reduced hypothetical bias but do not completely eliminated it |
| de-Magistris and Pascucci (2014) | Solemn oath | Consumer food choice | Between | Relative | Yes | - MWTP estimates were statistically lower with the oath script than without the oath script. |
| Carlsson et al. (2017) | Solemn oath | Commuter choice | Between | Relative | No | - Commuters' monetary trade-offs were the same regardless of the oath script |
| Kemper et al. (2020) | Solemn oath | Poultry products | Between | Relative | Yes | - Honesty oath reduced WTP<br>- Honesty oath influenced respondents thought processes |
| Lin et al. (2017) | Solemn oath + Honesty priming + Cheap talk | Food choice | Between | Relative | Mixed | - No significant differences in WTP values for between the various mitigation methods and a control group<br>- HB effect was likely not significant |
| Mamkhezri et al. (2020) | Solemn oath | Solar energy plans | Between | Relative | No | - Similar WTP estimates obtained across two treatments: with and without oath scripts |
| Ladenburg and Olsen (2014) | Opt-out/budget reminder + Cheap talk | Public good (Urban project) | Between | Relative | Yes | - Opt-out reminder significantly reduced total WTP and to some extent also MWTP beyond the capability of the cheap talk alone<br>- Introducing opt-out reminders as a supplement to a short CT script reduced welfare measures at the decision-to-opt-in level but not at the MWTP level |
| Varela et al. (2014) | Opt-out/budget reminder + Cheap talk | Forest fire prevention program | Between | Relative | Mixed | - The inclusion of a single opt-out reminder did not sufficiently improve the cheap talk effect |
| Alemu and Olsen (2018) | Opt-out/budget reminder | Novel food products | Between | Absolute | Yes | - HB effect was significant<br>- Repeated opt-out reminder mitigated hypothetical bias differently across different attributes |
| Penn et al. (2019) | Opt-out/budget reminder | Visits to Hawaiian beaches | Between | Relative | Yes | - Individual WTP differed between forced and unforced choice sets<br>- Evidence supports the use of unforced choice designs |
| Gschwandtner and Burton (2020) | Opt-out/budget reminder + Honesty priming + Cheap talk | Organic food product | Between | Relative | Yes | - A budget reminder combined with cheap talk script appeared to have reduced hypothetical bias more successfully than honesty priming |
| Cook et al. (2007) | Time to think | Medical treatment | Between | Relative | Yes | - Respondents who were given an overnight time to think made fewer choices that violated internal validity<br>- Respondents who were given time to think had lower average WTP |
| Cook et al. (2012) | Time to think | Medical treatment | Between | Relative | Yes | Average WTP dropped approximately 40% when respondents were given an overnight time to think |
| Ozdemir (2015) | Time to think | Medical treatments | Between | Relative | Yes | - Time-to-think approach has the potential to increase data validity |
| | | | | | | Possible drawbacks are increase in costs and strategic behaviour, and decrease in response rate |
| Tilley et al. (2016) | Time to think | Public hygiene intervention program | Between | Absolute | No | - Significant differences found in the choice behaviour of the subsamples<br>- The stated WTA estimates were far below those revealed by actual behaviour for both subsamples |
| Ben-Akiva and Morikawa (1990) | Pooled estimation with RP | Commuter mode choice | Within | Absolute | Yes | If properly corrected for biases, SP data could have predictive validity. |
| Hensher and Bradley (1993) | Pooled estimation with RP | Commuter mode choice | Within | Absolute | Yes | The signs and statistical significance of the estimated parameters in the jointly estimated model were more consistent with the RP model compared to the SP-only model |
| Adamowicz et al. (1994) | Pooled estimation with RP | Recreational site choice | Within | Absolute | Yes | Combined model yields less biased estimates |
| Brownstone et al. (2000) | Pooled estimation with RP | Vehicle purchase | Within | Absolute | Yes | Joint estimation improved accuracy of parameter estimates and market predictions |
| Duann and Shiaw (2001) | Pooled estimation with RP | Commute mode choice | Within | Absolute | Yes | SP only model underestimated VOT which was corrected in a joint estimation |
| Mark and Swait (2004) | Pooled estimation with RP | Physician's choice of prescription | Within | Absolute | Yes | Joint estimation can correct utility scales associated with SP data |
| Börjesson (2008) | Pooled estimation with RP | Commuter trip trimming choices | Within | Absolute | Yes | Systematic differences between SP and RP data were mitigated by joint estimation |
| Resano-Ezcaray et al. (2010) | Pooled estimation with RP | Food choice | Within | Absolute | Yes | Joint estimation improved market share prediction |
| Brooks and Lusk (2010) | Pooled estimation with RP | Food choice | Within | Absolute | Yes | Joint estimation improved accuracy of parameter estimates |
| Whitehead and Lew (2019) | Pooled estimation with RP | Recreational site choice | Within | Absolute | Yes | Differences between RP and SP estimates of total number of trips were mitigated by joint estimation |
| Buckell and Hess (2019) | Pooled estimation with RP | Tobacco demand | Within | Absolute | Yes | Embedding RP data in the model maked a substantial difference to the forecasts |
| Hultkrantz and Savsin (2017) | Referencing & realistic design | Value of time | Between | Relative | Mixed | - Significant differences found in the choice behaviour of the subsamples<br>- Referencing does affect responses by reducing the elicited implicit VoT<br>- Assuming that the SP-VoT is biased downwards, the bias would be further magnified by the referencing design |
| Chiu and Guevara (2019) | Referencing & realistic design | Commuter mode choice | Within | Absolute | Yes | - Downward hypothetical bias in SP-based VoT estimates was substantial but it was eliminated in the SP-off-RP setting<br>- Individual RP observations were better predicted using SP-off-RP than SP, hence further evidence of bias mitigation |
| Vossler et al. (2012) | Perceived consequentiality, real talk and | Tree planting project | Between | Absolute | Yes | - SP WTP estimates matched the financially binding incentive compatible treatment only for participants who believed their responses had more than a weak chance of influencing policy |
| | consequentiality script | | | | | |
| Interis and Petrolia (2014) | Perceived consequentiality, real talk and consequentiality script | Land restoration project | Between | Relative | Mixed | - Respondent perception of consequentiality was only effective in the multinomial choice and not the binary choice |
| Lewis et al. (2016) | Perceived consequentiality, real talk and consequentiality script | Food choice | Between | Relative | Yes | - Subjects who saw the consequentiality script had higher belief their responses will be consequential<br>- Consequentiality script decreased likelihood of choosing none-of-these options |
| Oehlmann and Meyerhoff (2017) | Perceived consequentiality, real talk and consequentiality script | Renewable energy | Between | Relative | Mixed | - Consequentiality script made subjects more inclined to perceive their responses are at least somewhat consequential<br>- WTP did not differ across treatments |
| Li et al. (2017) | Perceived consequentiality, real talk and consequentiality script | Food choice | Between | Relative | Mixed | - Belief in consequentiality increased WTP |
| Zawojska et al. (2019a) | Perceived consequentiality, real talk and consequentiality script | Renewable energy | Between | Relative | Mixed | - Policy (payment) consequentiality decreased (increased) cost sensitivity |